\documentclass[sigplan,nonacm,screen, bookmarksnumbered,unicode,urlbreakonhyphens=false]{acmart}

\usepackage{xcolor}
\usepackage{xurl}

\definecolor{mylinkblue}{HTML}{0000EE} % classic link blue

\AtBeginDocument{\hypersetup{
  colorlinks=true,
  linkcolor=mylinkblue,   % refs to sections/figures
  citecolor=mylinkblue,   % \cite
  urlcolor=mylinkblue,    % \url and \href{http://...}
  anchorcolor=mylinkblue
}}

\AtBeginDocument{%
  }

\acmISBN{978-1-4503-XXXX-X/2018/06} % change based on email

\usepackage{nicematrix}
\usepackage[utf8]{inputenc}

\newcommand{\etal}{{\it et al.}}
\newcommand{\tr}{\mathrm{Tr}}

\usepackage{bbold}

\usepackage{tikz}
\usetikzlibrary{quantikz2}
\usetikzlibrary{topaths,calc}
\usepackage{float}
\usepackage{amsmath}
\usepackage{ulem}
\usepackage{braket}
\usepackage{todonotes}
\usepackage{graphicx}
\usepackage{subcaption}
\usepackage{mathtools}

\usepackage[english]{babel}
\usepackage{amsthm}
\usepackage{xspace}

\usepackage{physics}
\usepackage{dsfont}

\usepackage{enumitem}

\def\arrvline{\hfil\kern\arraycolsep\vline\kern-\arraycolsep\hfilneg}

\newcommand{\cswap}{{CSWAP}}
\newcommand\name{{\sc COMPAS}}

\copyrightyear{2026}
\acmYear{2026}
\setcopyright{cc}
\setcctype{by}
\acmConference[ASPLOS '26]{Proceedings of the 31st ACM International Conference on Architectural Support for Programming Languages and Operating Systems, Volume 2}{March 22--26, 2026}{Pittsburgh, PA, USA}
\acmBooktitle{Proceedings of the 31st ACM International Conference on Architectural Support for Programming Languages and Operating Systems, Volume 2 (ASPLOS '26), March 22--26, 2026, Pittsburgh, PA, USA}
\acmPrice{}
\acmDOI{10.1145/3779212.3790143}
\acmISBN{979-8-4007-2359-9/2026/03}

\begin{document}
\title{\name: A Distributed Multi-Party SWAP Test for Parallel Quantum Algorithms}

\author{Brayden Goldstein-Gelb}
\authornote{Both authors contributed equally to this work.}
\email{brayden_goldstein-gelb@alumni.brown.edu}
\affiliation{%
  \institution{Brown University}
  \city{Providence}
  \state{Rhode Island}
  \country{USA}
}

\author{Kun Liu}
\authornotemark[1]
\email{kun.liu.kl944@yale.edu}
\affiliation{%
  \institution{Yale University}
  \city{New Haven}
  \state{Connecticut}
  \country{USA}
}

\author{John M. Martyn}
\email{john_martyn@fas.harvard.edu}
\affiliation{%
  \institution{Pacific Northwest National Lab}
  \city{Richland}
  \state{Washington}
  \country{USA}
}
\affiliation{%
  \institution{Harvard University}
  \city{Cambridge}
  \state{Massachusetts}
  \country{USA}
}

\author{Hengyun (Harry) Zhou}
\email{hyzhou@quera.com}
\affiliation{%
  \institution{QuEra Computing Inc.}
  \city{Boston}
  \state{Massachusetts}
  \country{USA}
}

\author{Yongshan Ding}
\email{yongshan.ding@yale.edu}
\affiliation{%
  \institution{Yale University}
  \city{New Haven}
  \state{Connecticut}
  \country{USA}
}

\author{Yuan Liu}
\email{q_yuanliu@ncsu.edu}
\affiliation{%
  \institution{North Carolina State University}
  \city{Raleigh}
  \state{North Carolina}
  \country{USA}
}

\renewcommand{\shortauthors}{Brayden Goldstein-Gelb et al.}

\newcommand{\asplossubmissionnumber}{2310} 



\begin{abstract}
    The limited number of qubits per chip remains a critical bottleneck in quantum computing, motivating the use of distributed architectures that interconnect multiple quantum processing units (QPUs). However, executing quantum algorithms across distributed systems requires careful co-design of algorithmic primitives and hardware architectures to manage circuit depth and entanglement overhead. We identify multivariate trace estimation as a key subroutine that is naturally suited for distribution, and broadly useful in tasks such as estimating R\'enyi entropies, virtual cooling and distillation, and certain applications of quantum signal processing. In this work, we introduce \name{}, an architecture that realizes multivariate trace estimation across a multi-party network of interconnected modular and distributed QPUs by leveraging pre-shared entangled Bell pairs as resources. \name{} adds only a constant depth overhead and consumes Bell pairs at a rate linear in circuit width, making it suitable for near-term hardware. Unlike other schemes, which must choose between asymptotic optimality in circuit depth or GHZ width, \name{} achieves both at once.
    Additionally, we analyze network-level errors and simulate the effects of circuit-level noise on the architecture.\footnote{The code is available at~\href{https://github.com/kunliu7/Distributed-Q-Algo}{github.com/kunliu7/Distributed-Q-Algo}.}
\end{abstract}


\received{19 August 2025}
\received[accepted]{4 January 2026}

\maketitle

\section{Introduction}
%\subsection{}

% Scalable quantum computers capable of solving classically intractable problems often require millions of qubits~\cite{gidney2021factor,gidney2025factor}. \JM{Are we keeping this sentence, or getting rid of it?}

%\JM{inclue a citation for this claim, e.g. https://arxiv.org/abs/1905.09749 or https://arxiv.org/abs/2505.15917} 
Scaling physical quantum systems while maintaining high qubit connectivity and gate fidelity presents significant challenges due to constraints in control complexity, wiring density, power bottlenecks, and maintenance overhead. Parallel and distributed quantum computing~\cite{margolus1990parallel, caleffi_2024_distributed, beals_efficient_2013} have emerged as two promising paradigms to address these limitations by leveraging modular designs that interconnect smaller quantum processors via quantum communication channels. 
% Efficiently executing generic quantum algorithms in such architectures requires sophisticated methods of distributed computing, 
%that balance circuit depth, qubit overhead, noise accumulation, and communication costs \cite{kim_fault-tolerant_2024,mor-ruiz_influence_2024,NoisyDistributedQC}.
However, harnessing such modular architectures demands careful co-design of quantum algorithms and the underlying hardware to overcome the fundamental tension between locality, latency, and fidelity~\cite{kim_fault-tolerant_2024,mor-ruiz_influence_2024,NoisyDistributedQC}. One fundamental challenge is how to perform efficient remote entangling operations across multiple parties. 
% Developing optimized distributed algorithms is crucial to harnessing the full computational power of modular quantum systems while overcoming hardware and architectural constraints. \YD{This paragraph can be made more concise: However, harnessing such modular architectures demands careful co-design of quantum algorithms and the underlying hardware to overcome the fundamental tension between locality, latency, and fidelity~\cite{kim_fault-tolerant_2024,mor-ruiz_influence_2024,NoisyDistributedQC}.}

% \YD{This paper presents \name{} (XXX full name), a co-designed hardware/software framework for executing multivariate trace estimation, a core quantum primitive, across distributed quantum systems. At the heart of \name{} is the key insight that the program structure of these algorithms permits a \textbf{structured decomposition of computation and communication}: local operations can be kept shallow, while remote gates can be efficiently handled using Bell-pair-assisted teleportation.} 
In distributed quantum computing, efficient execution of algorithms relies on high-rate and low-latency communication between modules \cite{parekh_quantum_2021,Cacciapuoti2020quantum}. The basic resource for such a modular system is a Bell pair shared between physical qubits in different nodes, enabling quantum-state teleportation and remote gate operations. Developing an interface with sufficient rate and fidelity to support high-fidelity computation remains a significant challenge. There has been tremendous experimental progress developing interconnections among quantum systems, such as using cryogenic microwave links for superconducting qubits or optical photons for atomic qubits \cite{ang2024arquin,main2025distributed}. Yet, state-of-the-art results for remote entanglement between neutral atom or trapped ion qubits only achieves rates of a few hundred Hz with entangling fidelities in the high 90\%~\cite{Stephenson_2020, jing2019entanglement, Ritter_2012, Hucul_2014, Pompili_2021}. Entanglement distillation protocols have also been developed to further boost the fidelity of entanglement while maintaining a high rate and low latency \cite{rozpkedek2018optimizing}. Despite this progress, architectures that can efficiently connect modular and distributed quantum hardware with generic high-level quantum algorithms are much underdeveloped. 

This paper introduces \name{} (\emph{Co}mpiling a \emph{M}ulti-Party SWAP Test for \emph{P}arallel \emph{A}lgorithms on Distributed Quantum \emph{S}ystems) as a co-designed hardware/software framework for executing multivariate trace estimation, a core quantum primitive, across distributed quantum systems. At the heart of \name{} is the key insight that the program structure of these algorithms permits a \textbf{structured decomposition of computation and communication}: local operations can be kept shallow, while remote gates can be efficiently handled using Bell pair assisted teleportation.  

More specifically, \name{} performs a particular multivariate trace estimation known as the \textit{multi-party SWAP test}, which underlies a wide range of parallel algorithms and protocols in quantum computing. Unlike general quantum circuits, the multi-party SWAP test has a highly structured design, which presents an opportunity for targeted compiler optimizations that minimize communication costs while preserving algorithmic running time and fidelity.
Distributed quantum circuits must simultaneously account for multiple conflicting constraints, including spatial limitations (qubit count), temporal efficiency (circuit depth), and inter-node communication (entanglement distribution and non-local operations). Logical circuits designed for monolithic architectures often do not translate efficiently to distributed systems, necessitating specialized compiler techniques to minimize long-range operations and reduce interconnect overhead. Without deep optimization, the resource overhead associated with parallel quantum algorithms can quickly become prohibitive, negating the theoretical advantages of parallel execution. A tailored compilation approach must consider qubit routing, gate scheduling, and efficient use of entanglement to ensure that logical operations can be executed with minimal additional cost while maintaining high fidelity.
\name{} provides an architecture design that naturally separates the individual pieces of data involved in the multi-party SWAP test into distinct modules. This reduces both the amount of network communication required, because state preparation can be performed within individual modules, and improves synchronization and parallelism, because the preparation of these inputs can happen in parallel across different modules.

In presenting \name{}, we develop techniques to adapt the Quek \etal{} implementation of the multi-party SWAP test \cite{quek_2024} into a distributed quantum algorithm and architecture, while maintaining constant circuit depth independent of the number of modules. The constant-depth property is crucial as it ensures that the runtime is not slowed down by communication between nodes, thereby enabling true parallelism at scale. 
Our approach optimizes both qubit usage and inter-node communication, while also remaining robust to noisy communication and gate errors, which we confirm via circuit-level simulation. 
We further highlight applications of \name{} to a wide range of computation tasks, including R\'enyi entropy calculation~\cite{Yirka_2021, Islam_2015}, entanglement spectroscopy~\cite{Johri_2017}, virtual cooling and distillation~\cite{Huggins_2021, Cotler_2019}, and parallel quantum signal processing~\cite{martyn_2024_parallel}.
In aggregate, by carefully balancing circuit design, entanglement distribution, and hardware constraints, we provide a pathway toward scalable, high-performance quantum computing on distributed architectures.

\subsection{Results and Paper Outline}

We begin with an overview of the necessary background in Sec.~\ref{sec:background}, reviewing methods in distributed quantum computing and the multi-party SWAP test. We then describe our proposed architecture in Sec.~\ref{sec:approach}. A central contribution of this section is a distributed implementation of the controlled-SWAP (\cswap{}) gate, which is a key operation used in recent multivariate trace estimation algorithms~\cite{quek_2024}. In Secs.~\ref{sec:telegate_design} and \ref{sec:teledata_design}, we present two such approaches—one based on the teledata primitive, and the other on the telegate primitive, the two core tools of distributed quantum computing.
In Sec.~\ref{sec:resources}, we analyze the resource costs of the proposed architectures, including circuit depth, and entanglement consumption.
We simulate and analyze the impact of gate-level noise in Sec.~\ref{sec:error_analysis}, and finally highlight applications of the proposed architecture to various quantum algorithms in Sec.~\ref{sec:applications}.

\section{Background}\label{sec:background}
\subsection{High-Level Overview}

A quantum state on $n$ qubits is described by a $2^n \times 2^n$ Hermitian matrix $\rho$, that is positive semi-definite and has unit trace, $\tr(\rho)=1$. Quantum states encode information about the physical system and its properties. As such, many relevant computational tasks require comparing many quantum states, for example in evaluating their overlaps or performing property testing.

The multi-party SWAP test is a standard tool for comparing many quantum states. It operates by extracting the trace of a product of states, a quantity in the interval $[0, 1]$ that provides information about the states' overlaps, spectra, and entanglement structure.
However, implementing the multi-party SWAP test on near-term hardware is expensive due to hardware limitations. In spite of this, distributed quantum computing offers a potential remedy by distributing computational tasks across a network of interconnected quantum processing units (QPUs). This reduces the computational load on each device, but can incur additional communication overhead.

In this work, we present a distributed implementation of the multi-party SWAP test that simultaneously achieves optimal latency and memory usage. To our knowledge, this is the first approach to do so.
Our construction, and distributed quantum protocols in general, use pairs of correlated (entangled) states known as Bell pairs, which are distributed between parties in advance.
These pre-shared resources enable spatially separated QPUs to interact with each other without directly transmitting quantum states during runtime.
We leverage this capability to distribute the SWAP test over many parties while consuming Bell pairs as a resource.

\subsection{Distributed Quantum Computing}
 
In this work, we distinguish between \textit{parallel} and \textit{distributed} quantum algorithms. We use parallel quantum algorithms to refer to algorithms that reduce circuit depth at the expense of increased qubit count. Such parallelism may occur on a single quantum processor or across multiple processors. 
 
\begin{figure}
\centering
\subfloat[\centering State teleportation]{
\begin{quantikz}[row sep = 0.5em, column sep = 0.3em]
    \lstick{$\ket{\varphi}$} & \ctrl{1} & \gate{H} & & \meter{} \wire[d][2]{c} \\
    \lstick[2]{$\ket{\Phi^+}$} & \targ{} & & \meter{} \wire[d][1]{c}\\
    & & & \gate{X}  & \gate{Z} &  &  \\
\end{quantikz} \label{fig:state-tele}
}
\subfloat[\centering Gate teleportation]{
\begin{quantikz}[row sep = 0.5em, column sep = 0.3em]
    \lstick{$\ket{\varphi}$} & \ctrl{1}  & & & \gate{Z} & \\
    \lstick[2]{$\ket{\Phi^+}$} & \targ{} & & \meter{} \wire[d][2]{c}\\
    & \ctrl{1} & \gate{H} & \push{\ }  & \meter{} \wire[u][2]{c}   \\
    \lstick{$\ket{\psi}$} & \targ{} & & \gate{X}  &
\end{quantikz} \label{fig:gate-tele}
}

\caption{Methods of distributed quantum computing from \cite{caleffi_2024_distributed, ferrari_2021_compiler}. (a) Teleports the state $\ket{\varphi}$ between parties whereas (b) applies a CNOT gate with control located on one QPU and target residing by the other.
}
\label{fig:TeledataTelegate}
\end{figure}
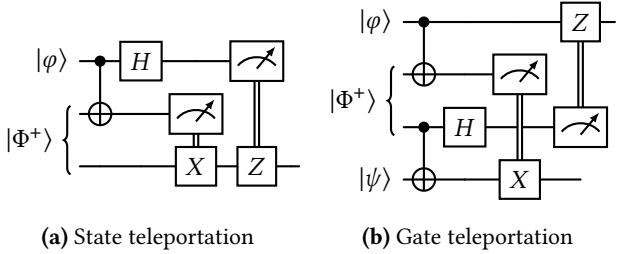

 In contrast, distributed quantum algorithms are designed to partition a computational task across multiple spatially separated QPUs, typically requiring classical or quantum communication between them. While having the potential to solve problems that exceed the resources of any single QPU \cite{Cuomo_2020}, distributed quantum computing faces obstacles not present in classical distributed computing. Notably, the no-cloning theorem forbids the copying of quantum states between processors. Instead, parties commonly interact through the \textit{teledata} and \textit{telegate} primitives (Fig. \ref{fig:TeledataTelegate}), which we collectively refer to as \textit{teleoperations}. Each teleoperation leverages pre-shared Bell pairs and classical communication as resources. A teledata operation refers to the teleportation of a quantum state between parties, whereas telegate operations allow one party to act as the control for a unitary acting on another. 

%Compiler techniques for cutting general quantum circuits into smaller sub-circuits and reconstructing via classical or quantum communication have been proposed but challenging...

\subsection{Multi-Party SWAP Test}\label{sec:multiparty_swap_test}
Many quantum algorithms and protocols involve extracting the trace of a product of quantum states. That is, given $k$ states $\{ \rho_i \}_{i=1}^k$, each acting on $n$ qubits, the goal is to compute the multivariate trace
\begin{equation}
    \tr(\rho_1 \rho_2 ... \rho_k) = \tr(\prod_{i=1}^k \rho_i). 
\end{equation}
This task arises in applications like calculating Rényi entropies, estimating eigenvalues of a quantum state, and mitigating gate errors on quantum devices.

A powerful method for estimating these quantities is the \textit{multi-party SWAP test}. 
Fundamentally, the multi-party SWAP test extends the standard SWAP test---which measures the overlap between quantum states~\cite{Buhrman_2001}---to higher-order products. This method uses the identity that the expectation value of a cyclic shift operator acting on a tensor product state 
\begin{equation}
    \rho_1 \otimes \rho_2 \otimes \cdots \otimes \rho_k = \bigotimes_{i=1}^k \rho_i
\end{equation}
equals the trace of the corresponding multiplicative product: 
\begin{equation} 
\tr\left( W_{\sigma} \cdot \bigotimes_{i=1}^k \rho_i \right) = \tr\left( \prod_{i=1}^k \rho_i \right), 
\end{equation} 
where $W_{\sigma}$ is the unitary corresponding to the cyclic permutation $\sigma = (1\ 2\ \ldots\ k)$. This operation maps index $i$ to $i + 1$ (mod $k$), and acts as 
\begin{equation} 
W_{\sigma} \left[ \rho_1 \otimes \rho_2 \otimes \cdots \otimes \rho_k \right] W_{\sigma}^\dagger = \rho_k \otimes \rho_1 \otimes \cdots \otimes \rho_{k-1}. 
\end{equation}

The multi-party SWAP test thus enables the computation of multiplicative products by evaluating $W_\sigma$ in the corresponding tensor product state. Crucially, this is achieved without sequentially multiplying the states $\rho_i$, but rather by arranging them in parallel and extracting an expectation value, effectively parallelizing their multiplication.

At the operational level, there are several approaches for performing the multi-party SWAP test. The simplest such method directly applies the Hadamard test to $W_\sigma$ and incurs a circuit depth $\mathcal{O}(k)$~\cite{Johri_2017, Yirka_2021}. More recent works have introduced implementations of the multi-party SWAP test that achieve constant circuit depth \cite{Suba__2019, quek_2024}. In \cite{quek_2024}, Quek \etal{} propose a constant depth method that first prepares an ancilla system in a special GHZ-like state, which is then used to measure the cyclic shift. It is this construction that will form the foundation of our proposed distributed architecture.

\begin{figure*}[ht]
  \centering
  \includegraphics[trim={0 1cm 0 0}, width=\textwidth]{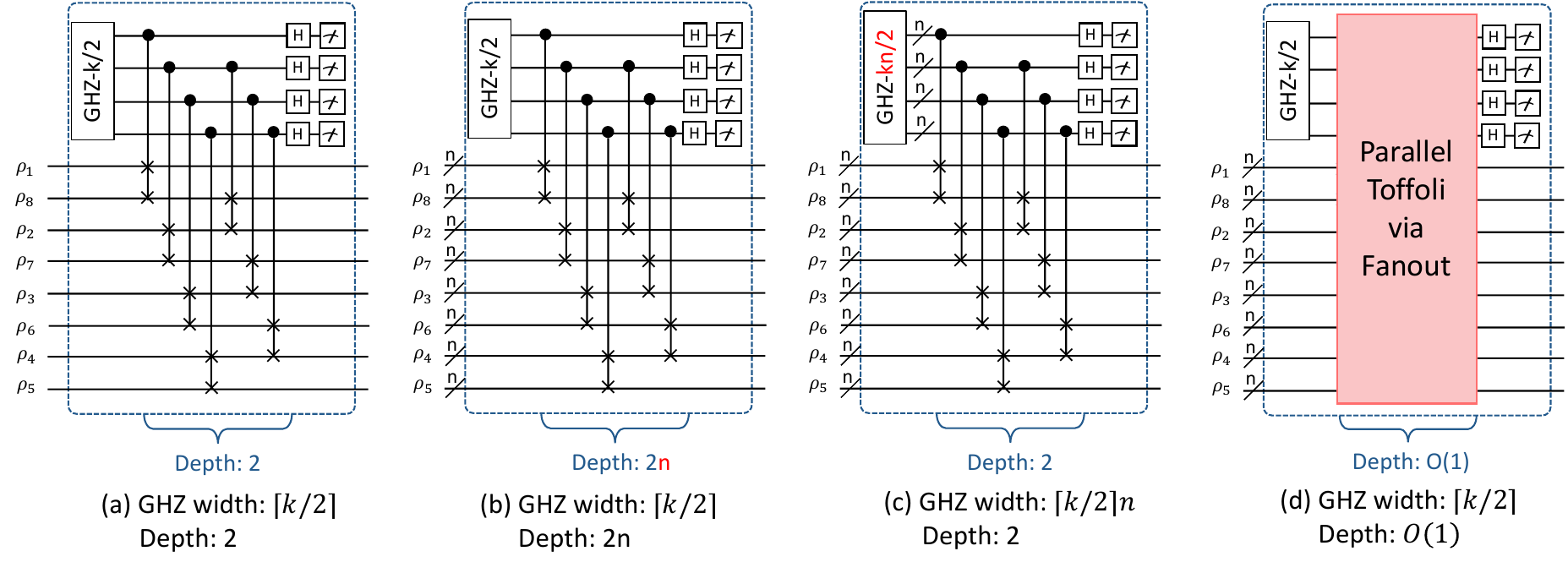}
  \caption{
  Comparing different implementations of $k$-party SWAP test.
  $n$: width of $\rho$.
  (a) $n=1,k=8$ example. GHZ width and circuit depth are $\lceil k/2 \rceil$ and $2$, respectively.
  For $n$-qubit state $k$-party SWAP test, ~\cite{quek_2024}
  propose either (b) by increasing the depth but keeping the width of GHZ as $\lceil k/2 \rceil$, or (c) by increasing width of GHZ to $\lceil k/2\rceil n$ but keeping the depth constant as $2$. (d) This work.
  We preserve
  GHZ width as $\lceil k/2\rceil$ and circuit depth as constant by applying parallel Toffoli via Fanout (Sec.~\ref{sec:parallel_toffoli}).
  }
  \label{fig:schemes}
\end{figure*}

In more detail,~\cite{quek_2024} realizes the multi-party SWAP test by the following steps, as shown in Fig.~\ref{fig:schemes}(a): 
\begin{enumerate}
    \item Prepare a $\lceil k/2 \rceil$-qubit GHZ state, serving as control qubits for the states $\rho_i$ permuted by $W_\sigma$.
    \item Perform two rounds of \cswap{}s on the states $\rho_i$.
    \item Measure each qubit of the GHZ state in the X basis.
\end{enumerate}
Quek \etal{} show that by repeating this process, one can estimate $\Im[\tr[\rho_1 \ldots \rho_k]]$. The real part, $\Re[\tr[\rho_1 \ldots \rho_k]]$ is obtained similarly, but with step 3 measured in the Y basis. Combined, these give an estimate of $\tr[\rho_1 \ldots \rho_k]$.

\subsection{Implementation for Multi-Qubit States}

% \KL{
In~\cite{quek_2024}, Quek \etal{} generalize the 
multi-party SWAP test from one-qubit states (Fig.~\ref{fig:schemes}a)
to multi-qubit states where \( n > 1 \),
by either using a deeper circuit or increasing the width of the GHZ state,
as shown in Figs.~\ref{fig:schemes}b and~\ref{fig:schemes}c, respectively.
In this work, we propose a new implementation of the multi-party SWAP test
for \( n \)-qubit states, 
which maintains both the optimal depth and width of the circuit,
as shown in Fig.~\ref{fig:schemes}d.
We achieve this by using parallel Toffoli gates via Fanout gates,
which will be introduced in Sec.~\ref{sec:parallel_toffoli}.
% }

\subsection{Naive Distributed Implementation}
\label{sec:distributing_rhos}
\begin{figure}[ht]
  \centering
  \includegraphics[trim={0 1cm 0 0}, width=\columnwidth]{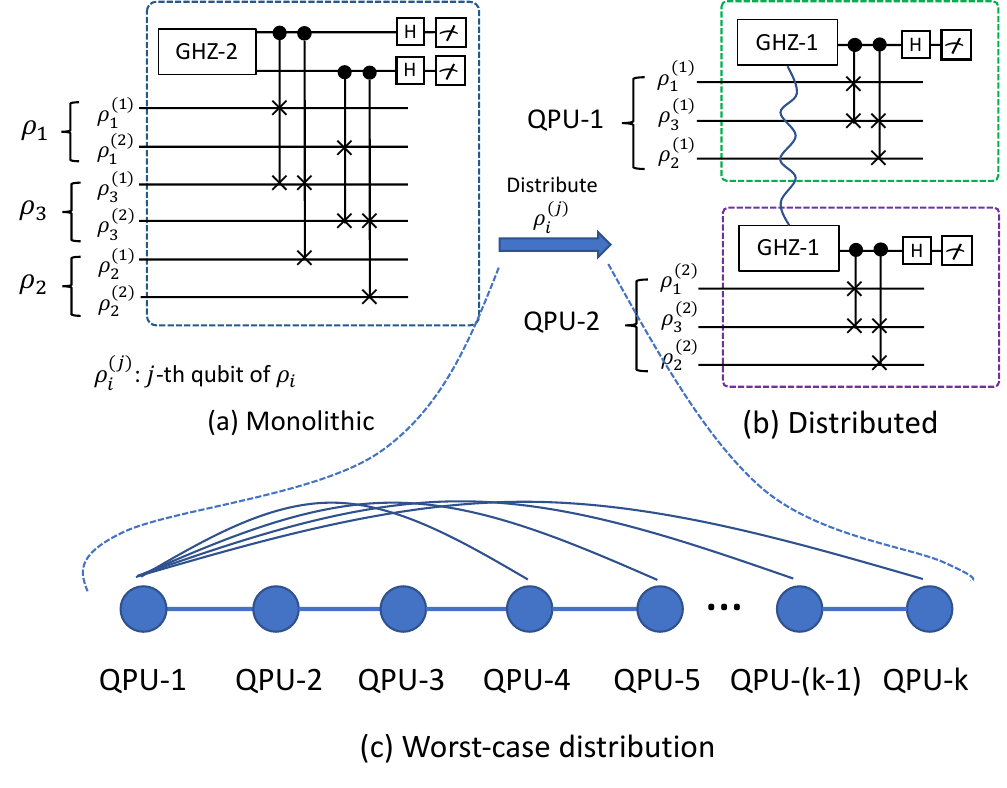}
  \caption{
   A depiction of the naive distribution approach for a $k=3$-party SWAP test for $n=2$-qubit states. (a) Execution on a single QPU. (b) Naive distributed implementation, where each state $\rho_i$ is partitioned into “slices’’ and each QPU is responsible for one slice. For a fixed $j$, all $\rho_i^{(j)}$ for $i \in [k]$ are sent to the same QPU, and the SWAP test is performed independently on each slice. In this example, QPU 1 is responsible for the first qubit of each state, QPU 2 the second, and QPU 3 (not shown) remains idle. (c) Worst-case distribution of states on a line topology, where QPU 1 resides at one endpoint  and must send its qubits to the other.
  }
  \label{fig:distribute_rho}
\end{figure}

% \KL{
Inspired by~\cite{liang_unified_2023}, we propose a naive implementation in which $\rho$ itself is distributed across $k$ QPUs. An example of this implementation with $k=3,n=2$ is shown in Fig.~\ref{fig:distribute_rho}(a,b).
To study this, let us denote the $j$-th qubit of $\rho_i$ as $\rho_i^{(j)}$.
We perform the following steps:
\begin{enumerate}
    \item For any fixed $j$, we distribute all the qubits $\rho_i^{(j)},\,\forall i\in [k]$ to the same QPU.
    \item Each QPU performs multi-party SWAP test locally on all the $j$-th qubits of the $\rho_i$'s collected in Step 1.
\end{enumerate}
% }

% \KL{
With $k$ QPUs, we can uniformly distribute $n$ tests across the QPUs,
resulting in an average of $n/k$ tests per QPU. In doing so, each QPU needs to distribute at most $n - n/k$ of its qubits.
We assume that the QPUs are connected in a line, which is the simplest connectivity. To teleport qubits from QPU $i$ to QPU $i+d$,
we first let each pair of nearest-neighbor QPUs to generate one Bell pair. We then use entanglement swapping to stitch these nearest-neighbor Bell pairs together and generate a long-range Bell pair between QPU $i$ and QPU $i+d$. This requires $d$ Bell pairs.
% }

% \KL{
The worst-case distribution is illustrated in Fig.~\ref{fig:distribute_rho}c,
where QPU 1 sits at the one endpoint of the line and needs to teleport qubits to the furthest QPUs.
Teleporting the qubits in this fasion requires $ n/k + (n/k+1) + ... + (n-1) = (n/k + n-1) \cdot (n - n/k)/2 = \mathcal{O}(n^2)$ Bell pairs in total.
If we assume that the computation continues after the multi-party SWAP tests, then we need to teleport each qubit back to its original QPU, and thus the Bell pair count doubles.
In this work, we will propose more efficient schemes, which require only a line topology and $\mathcal{O}(n)$ Bell pairs, as we  next introduce.

% }

\section{Approach}\label{sec:approach}
\subsection{Outline}

Our proposed \name{} architecture implements multivariate trace estimation by adapting the aforementioned approach of~\cite{quek_2024} to the distributed setting. We first outline how to use telegates to prepare the shared GHZ state of step (1). We then propose two methods of performing the \cswap{} subroutine between two parties, which is necessary for step (2). These modifications are thus sufficient to perform the multi-party SWAP test---and therefore the algorithms built upon it---in a fully distributed fashion.

\name{} maintains constant depth scaling, but introduces an additional cost of Bell pairs that must be shared between parties in advance. Given $k$ parties, each of width $n$, the number of Bell pairs scales asymptotically as $O(nk)$, for both methods of performing the \cswap{} subroutine. Below, we analyze the exact cost of each method and compare them.

\subsection{Architecture and Algorithm}\label{sec:architecture}

\begin{figure}
    \centering
\includegraphics[trim={0 0.5cm 0 0}, width=0.8\columnwidth]{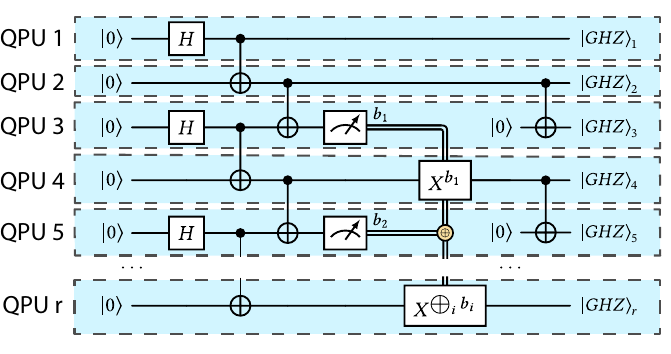}
% \begin{quantikz}[row sep=0.8em, column sep=1.2em]
%         \lstick{$\ket{0}_1$} & \gate{H} & \ctrl{1} &  &  &  & & & \\
%         \lstick{$\ket{0}_2$}&  & \targ{} & \ctrl{1} &  &  & &\ctrl{1}& \\
%         \lstick{$\ket{0}_3$}& \gate{H} & \ctrl{1} & \targ{} & \meter{} & \cwbend{1} \wire[l][1]["b_1"{above,pos=0.2}]{c} \setwiretype{n} & \lstick{$\ket{0}_3$} & \targ{} \setwiretype{q}  &\\
%         \lstick{$\ket{0}_4$}&  & \targ{} & \ctrl{1} &  & \gate{X^{b_1}}  & & \ctrl{1}& \\
%         \lstick{$\ket{0}_5$}& \gate{H} & \ctrl{0} & \targ{} & \meter{}  & \cwbend{-1} \wire[l][1]["b_2"{above,pos=0.2}]{c} \setwiretype{n} & \lstick{$\ket{0}_5$} & \targ{} \setwiretype{q} & \\
%         \setwiretype{n} \ldots &&&&&& \ldots \\
%         \lstick{$\ket{0}_r$} &  & \targ{}  &  &  & \gate{X^{\bigoplus_i b_i}}  & & \targ{} & \\
%     \end{quantikz}

\caption{Preparation of an $r$-party GHZ State in constant depth using gate teleportation adapted from~\cite{quek_2024}.}
    \label{fig:GHZPrep}
\end{figure} 
\begin{figure}
    \centering

% \begin{quantikz}[row sep=1.5em, column sep=1.5em]
%     \lstick[2]{
%         $\ket{GHZ}_1$\\
%         $\rho_1$
%     } &  & \ctrl{1} &  & \gate{H} & \meter{} \\
%      & \qwbundle{n} & \swap{1} & &  & \\
%     \lstick{$\rho_k$} &  \qwbundle{n} & \targX{} & \swap{2} &  &  \\
%     \lstick[2]{
%         $\ket{GHZ}_2$\\
%         $\rho_2$
%     } &  & \ctrl{1} &  \ctrl{0}& \gate{H} & \meter{} \\
%     & \qwbundle{n} & \swap{1} & \targX{}&  & \\
%     \lstick{$\rho_{k-1}$} & \qwbundle{n}  & \targX{} & \targX{} &  &  \\
%     \setwiretype{n} \cdots & & \cdots \\
%     \lstick[2]{
%         $\ket{GHZ}_{\lfloor \frac{k}{2} \rfloor}$\\
%         $\rho_{\lfloor \frac{k}{2} \rfloor}$
%     } &  & \ctrl{1} & \ctrl{1} & \gate{H} & \meter{} \\
%     & \qwbundle{n} & \swap{1} &\targX{} &  & \\
%     \lstick{$\rho_{\lceil \frac{k}{2} \rceil + 1}$} & \qwbundle{n}  & \targX{} &  \swap{2}& & \\
%     \lstick[2]{
%         $\ket{GHZ}_{\lceil \frac{k}{2} \rceil}$\\
%         $\rho_{\lceil \frac{k}{2} \rceil}$
%     } &  &  & \ctrl{0} & \gate{H} & \meter{} \\
%     & \qwbundle{n} &  &\targX{} &  & \\
% \end{quantikz}

\includegraphics[trim={0 0.5cm 0 0}, width=0.78\columnwidth]{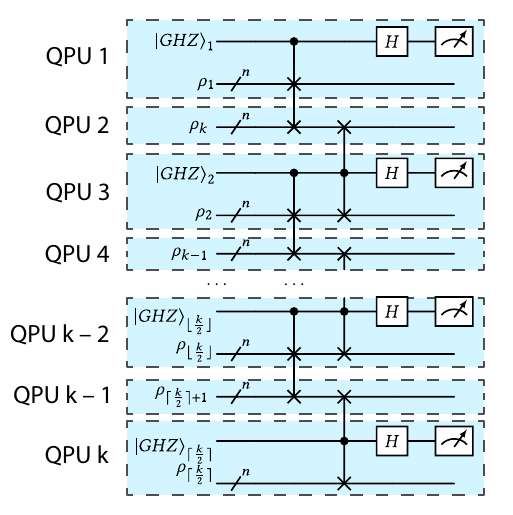}

\caption{Quantum circuit diagram for applying the $W_\sigma$ gate to the state $\bigotimes_{i=1}^k \rho_i$. The grouped states belong to the party. This figure is for the case when $k$ is odd. The even case is the same but without the bottom two wires.}
    \label{fig:PermutationCircuit}
\end{figure} 

We now adapt the construction outlined in \cite{quek_2024} to the distributed setting. As discussed in Sec.~\ref{sec:multiparty_swap_test}, the algorithm is primarily comprised of preparing a GHZ state, and performing a controlled cyclic-shift, and a final measurement. We first take note of the structure of the cyclic-shift operation, which involves two rounds of CSWAPs on the states $\rho_i$. As can be seen in Fig.\ref{fig:PermutationCircuit}, each of the $k$ states interacts with at most two others. By arranging the states in a interleaved $1, k, 2, k - 1, \ldots$ pattern, interactions are limited to adjacent neighbors in this ordering. We further observe that in step (2), each qubit of the GHZ state acts on a state during both rounds of \cswap{}s: first a \cswap{} with its right-hand neighbor, then a \cswap{} with its left-hand neighbor in the interleaved order.

This naturally lends itself to distributed network design with $k$ QPUs, where each QPU stores a single state and, if needed, its corresponding GHZ qubit. We illustrate this design in Fig.~\ref{fig:PermutationCircuit}.
Such a design has the advantage that communication between QPUs is only required during the multi-party SWAP test, and not during the preparation of $\rho$, thus alleviating communication requirements.

We next need to prepare a GHZ state with each qubit located on a different QPU. Toward this end, we modify the constant-depth construction used in~\cite{quek_2024} by replacing inter-QPU CNOT gates with their telegate counterpart. This circuit can be seen in Fig.~\ref{fig:GHZPrep}. We also note that where construction in~\cite{quek_2024} uses a $\lfloor k/2\rfloor$-party GHZ state, we instead require $\lceil k/2\rceil$ parties. This is because we do not reuse the first qubit of the GHZ state in order to prioritize local operations in the cyclic-shift stage.

It then remains to implement a two-party version of the \cswap{} operation used in step (2), while maintaining constant depth. A controlled-SWAP operation, as the name implies, swaps two $n$-qubit states, $\rho_i$ and $\rho_j$, conditional on a control qubit $\ket{\varphi}$. In the two-party setting, these states are distributed across two parties: Alice and Bob. Specifically, Alice stores both $\ket{\varphi}$ and $\rho_i$ on her QPU, while Bob stores $\rho_j$ on his. Alice and Bob are assumed to share Bell pairs in advance and may communicate via a classical channel. We present two constructions of such a two-party \cswap{} in Secs.~\ref{sec:telegate} and ~\ref{sec:teledata}. The two methods have very similar error rates and resource requirements. Though, due to its potentially lower memory footprint, we recommend the teledata construction, as stated in Sec. \ref{sec:resources}.

\begin{figure*}
    \centering
    \includegraphics[trim={0 0.5cm 0 0}, scale=0.75]{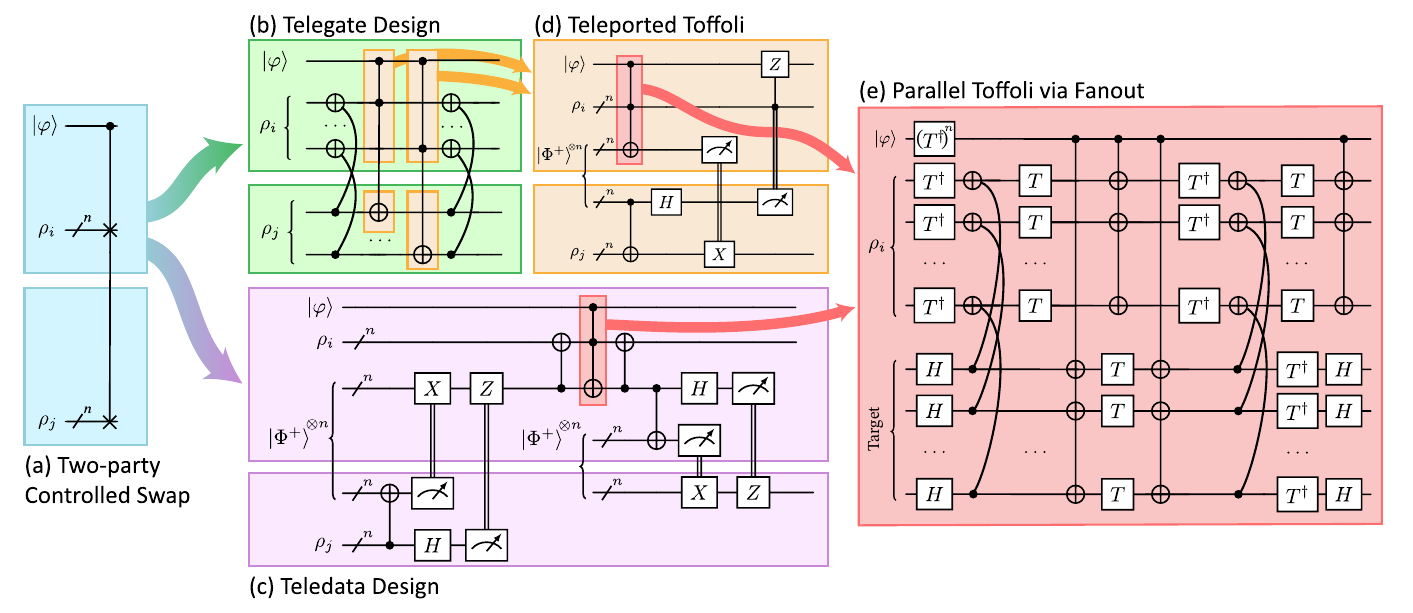}
    % \vspace{-0.3in}
    \caption{
    Decomposition of a two-party controlled-swap (a) via either the \textit{Telegate Design} (b) shown in green and described in Sec.~\ref{sec:telegate}, or the \textit{Teledata Design} (c) shown in purple and described in Sec.~\ref{sec:teledata}. Each arrow corresponds to unraveling the boxed gate at the source of the arrow into an equivalent circuit shown in the box to at the target of the arrow. 
    % The trash icons in the Teledata Design indicate that the corresponding qubit may be discarded and reset.
    }
    \label{fig:cswap_decomposition}

    % \vspace{-0.8em}
\end{figure*}
\subsection{Telegate Design}\label{sec:telegate}
\label{sec:telegate_design}
To build intuition for the telegate design, we begin by first considering the special case where $\ket{\varphi}, \rho_i$, and $\rho_j$ are all single-qubit states stored on the same QPU. In this scenario, the \cswap{} operation can be implemented using the following sequence of gates: 
\begin{enumerate}
    \item Perform a CNOT gate with  $\rho_j$ as the control qubit and $\rho_i$ as the target qubit.
    \item Perform a Toffoli gate with $\ket{\varphi}$ as the control qubit and $\rho_i, \rho_j$ as the target qubits.
    \item Perform another CNOT gate, again with  $\rho_j$ as the control qubit and $\rho_i$ as the target qubit.
\end{enumerate}

The $n$-qubit case can be achieved analogously. In steps (1) and (3), we perform $n$ CNOT gates in parallel, where the control qubit of each CNOT is the $l$th qubit of $\rho_j$, and the target qubit is the $l$th qubit of $\rho_i$ for $l = 1, \ldots, n$. Similarly, in step (2), we replace the single Toffoli gate with $n$ Toffoli gates, where the control qubits of each Toffoli are $\ket{\varphi}$ and the $l$th qubit of $\rho_i$, and the target qubit is the $l$th qubit of $\rho_j$ for $l = 1, \ldots, n$. These Toffoli gates can be executed in parallel using the \textit{Fanout method} described in Sec.~\ref{sec:parallel_toffoli}. 

Finally, we use telegates to accommodate the distributed setting. Specifically, we replace the CNOT gates previously described with teleported CNOT gates from ~\cite{caleffi_2024_distributed, ferrari_2021_compiler}, as shown in Fig.~\ref{fig:gate-tele}. Additionally, the $n$ Toffoli gates are replaced with teleported Toffoli gates modified from \cite{eisert_2000_optimal} (boxed in orange in Fig.~\ref{fig:cswap_decomposition}b and ~\ref{fig:cswap_decomposition}c). These teleported Toffoli gates can also be implemented in parallel using the \textit{Fanout method} for the local Toffoli gates (boxed in red in Fig.~\ref{fig:cswap_decomposition}d and ~\ref{fig:cswap_decomposition}e) which are used as part of the teleportation.

\subsection{Teledata Design}\label{sec:teledata}
\label{sec:teledata_design}
In the teledata design, the two-party \cswap{} is implemented by teleporting Bob's state to Alice's QPU, performing the \cswap~ locally, and teleporting his state back. In more detail:

\begin{enumerate}
    \item Bob teleports the $l$th qubit of $\rho_j$ to Alice's $l$th ancilla qubit in parallel for $l = 1, \ldots, n$.
    \item Bob resets the $n$ qubits used in the teleportation to $|0\rangle^{\otimes n}$.
    \item Alice performs a local \cswap, where $\ket{\varphi}$ is the control qubit and the states being swapped are $\rho_i$ and the $n$ ancilla qubits which now contain $\rho_j$. This local \cswap{} can be performed as described in Sec.~\ref{sec:telegate_design}.
    \item Alice teleports the $l$th ancilla qubit back to the Bob's $l$th qubit in parallel for $l = 1, \ldots, n$.
\end{enumerate}

Because teleportation preserves quantum state, any local operations applied to Bob's state while it resides on Alice's QPU are logically equivalent to performing them while the state resided on Bob's QPU. Thus, the above procedure correctly implements a two-party \cswap{} as desired.

\subsection{Parallel Toffoli via Fanout}
\label{sec:parallel_toffoli}

In both the telegate design in Sec.~\ref{sec:telegate} and the teledata design 
in Sec.~\ref{sec:teledata}, 
\( n \) Toffoli gates share a single 
control qubit \( \ket{\varphi} \), as shown in orange boxes in Fig.~\ref{fig:cswap_decomposition}b
and red boxes in Fig.~\ref{fig:cswap_decomposition}c and ~\ref{fig:cswap_decomposition}d.
Without optimization, these Toffoli gates must be executed sequentially, 
resulting in \( \mathcal{O}(n) \) depth.
In this section, we present a method to parallelize these shared-control Toffoli gates
based on~\cite{gokhale_quantum_2020}, achieving \( \mathcal{O}(1) \) depth.

\begin{figure}[t]
  \centering
  \includegraphics[trim={0 1cm 0 0}, width=\columnwidth]{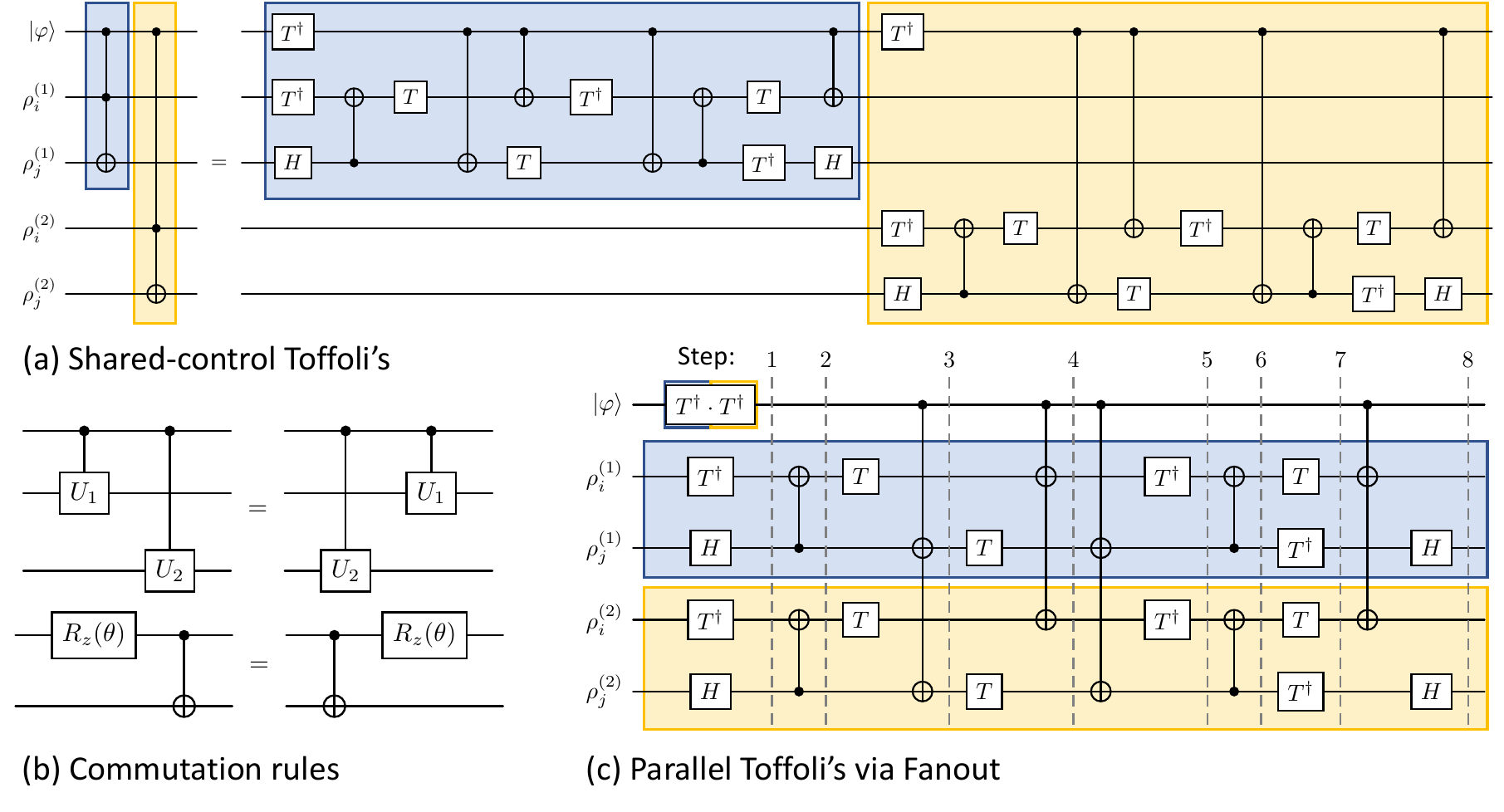}
  % \vspace{-0.3in}
  \caption{(a) Circuit diagram showing two shared-control Toffoli gates, where \( \rho_i^{(k)} \)
  denotes the \( k \)-th qubit of state \( \rho_i \).
  Both Toffoli gates utilize the optimal decomposition from~\cite{amy_meet---middle_2013}'s Fig.~7a.
  (b) Commutation rules enabling backward propagation of CNOTs and single-qubit gates
  from the second Toffoli gate~\cite{gokhale_quantum_2020}. 
  Adjacent CNOTs with shared control qubits combine to form Fanout gates.
  (c) Resulting parallel implementation using Fanout gates, with execution steps indicated by gray dashed lines.
  }
  \label{fig:parallel_toffoli}
\end{figure}

We begin by finding an optimal decomposition of a single Toffoli gate.
% The most efficient decomposition known to date
It was proposed in Fig.~7a of ~\cite{amy_meet---middle_2013},
requiring 7 $T$ gates with a \( T \)-depth of 4, and a total depth of 8.
We first place two shared-control Toffoli gates in the circuit,
as shown in Fig.~\ref{fig:parallel_toffoli}a.
Then, using the commutation rules in Fig.~\ref{fig:parallel_toffoli}b,
we push backward the CNOTs and single-qubit gates from the second Toffoli.
When two CNOTs share the same control qubit, they form a Fanout gate.
The resulting circuit is shown in Fig.~\ref{fig:parallel_toffoli}c,
maintaining the optimal \( T \)-depth of 4 and total depth of 8.
Crucially, both depths remain constant regardless of the number of qubits in \( \rho_i \).
% This implementation requires 4 Fanout gates.

\begin{figure}[ht]
  \centering
  \includegraphics[trim={0 1cm 0 0}, width=0.6\columnwidth]{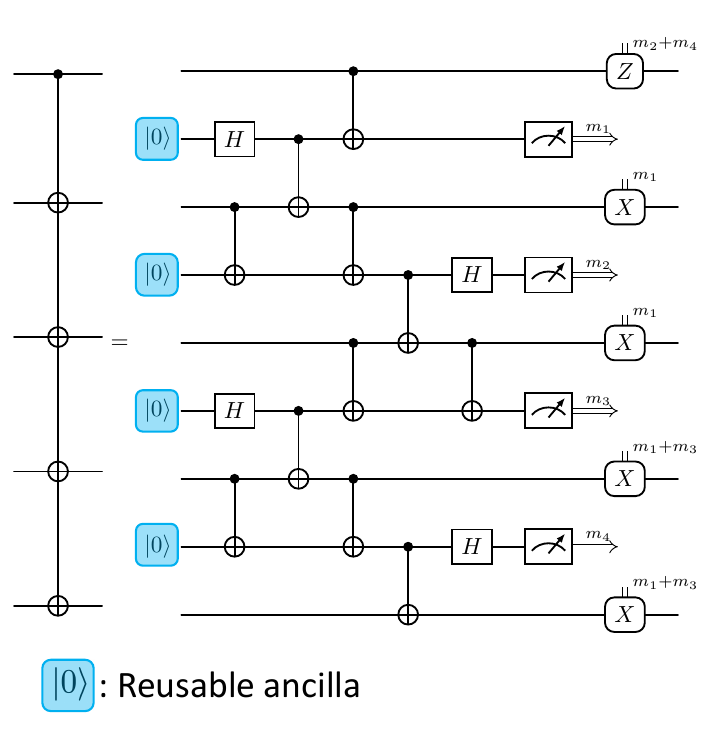}
  % \vspace{-0.1in}
  \caption{Implementation of a constant-depth Fanout gate with 4 target qubits~\cite{pham_2d_2013}.
  The circuit requires one ancilla qubit (blue boxes in the figure) initialized to \( \ket{0} \) per target qubit,
  and the ancilla qubits are reused across multiple Fanout gates.
  % The total circuit depth is 7, comprising measurement operations and subsequent Pauli corrections.
  }
  \label{fig:fanout}
  % \vspace{-0.4in}
\end{figure}

For implementing the Fanout gates, we employ the constant-depth circuit 
from~\cite{pham_2d_2013},
illustrated in Fig.~\ref{fig:fanout}.
This implementation requires one ancilla qubit initialized to \( \ket{0} \) for each target qubit
and achieves a depth of 7, including measurements and Pauli corrections.
% We observe that the two Fanout gates before and after
% the dashed line in Step 4 of Fig.~\ref{fig:parallel_toffoli}c
% can be combined.
By grouping qubits of the same state together
and optimizing single-qubit gate placement for clarity,
we obtain the final circuit shown in Fig.~\ref{fig:cswap_decomposition}e.
% This optimization reduces the number of Fanout gates to 3.
% Although one Fanout gate now targets \( 2n \) qubits
% with a depth of 9,
% the constant-depth nature of the Fanout operation ensures
% that the total circuit depth is reduced.

\subsection{Qubit Reuse}
\label{sec:qubit_reuse}
We assume that the ancilla qubits are reused across different rounds or steps,
which is a common technique in compiling quantum circuits 
with mid-circuit measurements and resets~\cite{decross_qubit-reuse_2023,hua_caqr_2023,fang_dynamic_2023,lin_reuse-aware_2024}.
For a Fanout gate with \( n \) target qubits,
we need \( n \) ancilla qubits initialized to \( \ket{0} \),
as shown in Fig.~\ref{fig:fanout}.
These ancilla qubits are reused across multiple Fanout gates,
i.e., all the Fanout gates in Fig.~\ref{fig:parallel_toffoli}c
share the same set of ancilla qubits; so 
\( n \) ancilla qubits are sufficient.
% However, for the circuit in Fig.~\ref{fig:cswap_decomposition}e,
% there are two Fanout gates targeting \( n \) qubits,
% but one Fanout gate targeting \( 2n \) qubits,
% so \( 2n \) ancilla qubits are required.
% Such an analysis is detailed in Sec.~\ref{sec:resources}.

% \vspace{-1\baselineskip}
\section{Resource Consumption}
\label{sec:resources}

\begin{figure}[t]
    % \centering

% \begingroup
% \setlength{\parskip}{0pt}   % no extra gap between blocks
% \setlength{\parindent}{0pt}
\begin{minipage}{0.48\textwidth}
        % \begin{table}[H]
            \centering
            \resizebox{\columnwidth}{!}{%
            \begin{tabular}{lllll}
            \hline
                 & Step                              & Ancilla   & Bell pairs & Depth       \\ \hline
            (a) Fig.~\ref{fig:GHZPrep}  & GHZ preparation                          & 1         & 2          & 9           \\ \hline
            (b1) Fig.~\ref{fig:cswap_decomposition}b & CNOT teleportation x2                     & 0         & 2n         & 3*2         \\
            (b2) Fig.~\ref{fig:cswap_decomposition}d & Toffoli teleportation                     & 0         & n          & 6       \\ \hline
            (b3) Fig.~\ref{fig:parallel_toffoli}c & Toffoli's, non-Fanout gates       & 0         & 0          & 4           \\
            (b4) Fig.~\ref{fig:parallel_toffoli}c & Toffoli's, Fanout gates x4        & n         & 0          & 7*4         \\ \hline
            (c)  & Readout                           & 0         & 0          & 2           \\ \hline
            (d)  & Total = (a) + (b1-b4) x2 + (c)    & n (reuse) & 2 + 3n*2   & 9+44*2+2=99 \\ \hline
            \end{tabular}%
            }
            \captionof{table}{
                Cost per QPU for the \emph{telegate} scheme (Sec.~\ref{sec:telegate}) using 4 Fanout gates
                (Fig.~\ref{fig:parallel_toffoli}c).
                Since we have two rounds of cSWAP gates in cyclic shift in
                multivariate trace estimation, we need to repeat (b1-b4).
                Ancilla can be reused across different rounds or steps.
            }
            \label{tab:telegate_cost_4fanouts}
        % \end{table}
    \end{minipage}
    \begin{minipage}{0.48\textwidth}
        % \begin{table}[H]
            \centering
            \resizebox{\columnwidth}{!}{%
            \begin{tabular}{lllll}
            \hline
                 & Step                           & Ancilla    & Bell pairs & Depth       \\ \hline
            (a) Fig.~\ref{fig:GHZPrep}  & GHZ preparation                      & 1          & 2          & 9           \\ \hline
            (b1) Fig.~\ref{fig:cswap_decomposition}c & Data teleportation                     & n          & 2n         & 8          \\ \hline
            (b2) Fig.~\ref{fig:parallel_toffoli}c & Toffoli's, non-Fanout gates    & 0          & 0          & 4           \\
            (b3) Fig.~\ref{fig:parallel_toffoli}c & Toffoli's, Fanout gates x4     & n          & 0          & 7*4         \\ \hline
            (c)  & Readout                        & 0          & 0          & 2           \\ \hline
            (d)  & Total = (a) + (b1-b3) x2 + (c) & 2n (reuse) & 2+2n*2     & 9+40*2+2=91 \\ \hline
            \end{tabular}%
            }
            \captionof{table}{
                Cost per QPU for the \emph{teledata} scheme (Sec.~\ref{sec:teledata}) using 4 Fanout gates
                (Fig.~\ref{fig:parallel_toffoli}c).
                Since we have two rounds of cSWAP gates in cyclic shift in
                multivariate trace estimation, we need to repeat (b1-b3).
                Ancilla can be reused across different rounds or steps.
            }
            \label{tab:teledata_cost_4fanouts}
        % \end{table}
    \end{minipage}
    \begin{minipage}{0.48\textwidth}
    % \captionof{table}{Cost per QPU for the telegate scheme ...}
    % \begin{table}[H]
    \centering
    \resizebox{\columnwidth}{!}{%
    \begin{tabular}{lllll}
    \hline
    Scheme                  & Ancilla & Bell pairs & Depth & Memory est. \\ \hline
    (a) Telegate, 4 Fanouts (Table~\ref{tab:telegate_cost_4fanouts}) & n       & 2 + 6n     & 99    & 19n + 6 \\
    % (b) Telegate, 3 Fanouts (Table~\ref{tab:telegate_cost_3fanouts}) & 2n      & 2 + 6n     & 85    & 20n + 6 \\
    \textbf{(b) Teledata, 4 Fanouts} (Table~\ref{tab:teledata_cost_4fanouts}) & \textbf{2n}      & \textbf{2 + 4n}     & \textbf{91}    & \textbf{14n + 6} \\
    % (d) Teledata, 3 Fanouts (Table~\ref{tab:teledata_cost_3fanouts}) & 3n      & 2 + 4n     & 81    & 15n + 6 \\
    (c) Naive Distribution (Sec.~\ref{sec:distributing_rhos}) & $n$      & $n(n+1)-\frac{n}{k}(\frac{n}{k}+1)$     & 76     & $\sim 3n^2$ \\
    % compare to teledata: 5 + 10/2 + 2 * (4 + 4*7) + 2
    \hline
    \end{tabular}%
    }
    \captionof{table}{
        Cost per QPU for the proposed algorithm.
        By distillation protocols
        in~\cite{pattison_fast_2024,ataides_constant-overhead_2025} where
        about 3 Bell pairs are required to distill 1 Bell pair,
        we estimate the memory by 3 times of the Bell pairs + ancilla.
        The bold font indicates the best scheme we recommend.
    }
    \label{tab:compare_cost}
% \end{table}
    \end{minipage}
\end{figure}
% \endgroup

In this section, we analyze the resource requirements of our proposed algorithm
in terms of ancilla qubits, Bell pairs, circuit depth, and memory consumption.
% We evaluate mainly two implementation schemes that combine either telegate or teledata approaches.
We first examine the cost of the telegate scheme in detail.
For the teledata scheme, we avoid redundancy by highlighting only its differences from the telegate scheme.
% We analyze the cost of telegate schemes using both 4 and 3 Fanout gates together,
% as they share the same table structure.
% For convenience, 
% we present the resource cost tables (Tables~\ref{tab:telegate_cost_4fanouts},
% \ref{tab:telegate_cost_3fanouts}, \ref{tab:teledata_cost_4fanouts},
% and \ref{tab:teledata_cost_3fanouts}),
% which shows the telegate and teledata circuits.
Note that we calculate the maximum cost per QPU,
as only QPUs that generate and maintain control qubits from the GHZ state
need to perform parallel Toffoli gates.
Other QPUs incur substantially lower costs.

\textbf{Telegate}, 
see Table~\ref{tab:telegate_cost_4fanouts}:
% \hzc{It seems like the output qubits are not used throughout the entire fanout, so the subsequent T gates in the circuit can be done at the same time as the rest of the fanout. As such, the total depth should be a bit smaller than just the sum.}
% \KL{But how do we deal with Pauli corrections in Fanout? E.g. before: $X^{m_1}$ first and T gate next on some data qubit. Now to do the T's at the same time of Fanout, we need to commute the T backward before $X^{m_1}$, is that correct?}
\textbf{(a)}: We require a GHZ state to serve as the control qubits for \cswap{}s, prepared using the circuit shown in Fig.~\ref{fig:GHZPrep}.
% QPU \( i \) requires 1 ancilla qubit and 2 Bell pairs to teleport the 2 CNOT gates 
% to QPU \( i-1 \) and QPU \( i+1 \).
% This process has a depth of 5 (including measurement and Pauli corrections),
% as shown in Table~\ref{tab:telegate_cost_4fanouts}a and Table~\ref{tab:telegate_cost_3fanouts}a.

\textbf{(b1)}: After GHZ state preparation, we can perform parallel \cswap{}s on
\( \rho_i \) and \( \rho_{j} \).
A \cswap{} gate decomposes into 2 CNOT gates and 1 Toffoli gate.
Since \( \rho_i,\rho_j \) have width \( n \),
we require $2n$ CNOTs and $n$ Toffoli gates,
as illustrated in Fig.~\ref{fig:cswap_decomposition}b.
The telegate scheme implements these CNOTs and Toffoli gates using gate teleportation.
For the $2n$ CNOTs, we employ gate teleportation in Fig.~\ref{fig:TeledataTelegate}b,
requiring $2n$ Bell pairs. 
% Since these $2n$ CNOT teleportations can occur in parallel,
% the depth is twice that of a single CNOT teleportation, i.e., $2\times3=6$,
% as shown in Table~\ref{tab:telegate_cost_4fanouts}b1 and Table~\ref{tab:telegate_cost_3fanouts}b1.

\textbf{(b2)}: For the $n$ shared-control Toffoli gates, we utilize the circuit in Fig.~\ref{fig:cswap_decomposition}d
for gate teleportation. In this circuit, all gates except
the shared-control Toffoli gates in the red box (which is a \( C^{n+1}X \) gate) can be executed in parallel.
% The depth values in Table~\ref{tab:telegate_cost_4fanouts}b2
% and Table~\ref{tab:telegate_cost_3fanouts}b2
% account for the gates during gate teleportation, excluding the shared-control Toffoli gates themselves.

\textbf{(b3-b4)}: For parallel Toffoli gates via Fanout gates,
we divide the circuit into two components: Fanout gates and non-Fanout gates.
% We present two approaches for parallelizing the Toffoli gates,
% using either 4 or 3 Fanout gates, as shown in Figs.~\ref{fig:parallel_toffoli}c
% and~\ref{fig:cswap_decomposition}e, respectively.
The Fanout gate, illustrated in Fig.~\ref{fig:fanout},
maintains a constant depth of \( 7 \) but requires \( n \) ancilla qubits,
where \( n \) represents the number of target qubits.
A key optimization here is qubit reuse, as described in Sec.~\ref{sec:qubit_reuse}.
Although we employ multiple Fanout gates, they can share the same set of ancilla qubits
as they operate at different time steps.
% All Fanout gates target \( n \) qubits,
% thus \( n \) ancilla qubits suffice, 
% as shown in Table~\ref{tab:telegate_cost_4fanouts}b4.
% In contrast, the 3 Fanout gates scheme requires two Fanout gates targeting \( n \) qubits and one targeting \( 2n \) qubits,
% necessitating \( 2n \) ancilla qubits
% as shown in Table~\ref{tab:telegate_cost_3fanouts}b4.
% Additionally, the non-Fanout gates 
% (bitwise CNOTs and T gates) in Fig.~\ref{fig:cswap_decomposition}e
% and Fig.~\ref{fig:parallel_toffoli}c
% can be executed in parallel with depths of 4 and 6 respectively, as shown in Table~\ref{tab:telegate_cost_4fanouts}b3
% and Table~\ref{tab:telegate_cost_3fanouts}b3.

\textbf{(d)}: For multivariate trace estimation, we perform two rounds of \cswap{} gates,
controlled by the same GHZ state.
This requires repeating steps (b1-b4) twice, doubling their Bell  cost and depth,
while steps (a) and (c) are performed only once.
Through qubit reuse, ancilla qubits are shared across different rounds,
resulting in a single count. 
% For example, in Table~\ref{tab:telegate_cost_4fanouts}d,
% we require \( n \) ancilla qubits, a Bell pair count $= (\text{a}) + (\text{b1} - \text{b4}) \times 2 + (\text{c}) = 2 + 2\times3n$,
% and a depth $= (\text{a}) + (\text{b1}-\text{b4}) \times 2 + (\text{c}) = 5 + 44\times2 + 2 = 95$.

\textbf{Teledata}, 
see Table~\ref{tab:teledata_cost_4fanouts}:
\textbf{(b1)}: Unlike the telegate scheme,
we directly teleport the data instead of decomposing \cswap{} gates into CNOTs and Toffoli gates.
Thus, the teledata scheme replaces (b1) CNOT teleportation and (b2) Toffoli teleportation
with (b1) data teleportation, while maintaining all other terms in the tables (Table~\ref{tab:teledata_cost_4fanouts}).
For teledata, we employ the circuit in Fig.~\ref{fig:cswap_decomposition}c, requiring a depth of 8 excluding the $n$ shared-control Toffoli gates (red box). 
The parallelization of these shared-control Toffoli gates matches that of the telegate schemes.

% \KL{
\textbf{Naive Distributed Implementation}: For Bell pair counts, refer to Sec.~\ref{sec:distributing_rhos}. After the distribution of $\rho_i$'s,
there is no further inter-QPU communication. Within each QPU, we can still apply parallel Toffoli via Fanout. Each QPU performs $n/k$ tests in parallel,
and each test will execute Fanout gates with $k$ target qubits, so $n$ ancilla qubits are required.
When computing the depth, we assume that all the Bell pairs are pre-generated,
regardless of their length.
% }

\textbf{Overall Recommendation}: The aggregate costs for each scheme are summarized in Table~\ref{tab:compare_cost}.
Given our focus on distributed QPU implementation, we must consider that Bell pairs typically require entanglement distillation due to noise.
Recent works~\cite{pattison_fast_2024,ataides_constant-overhead_2025} have demonstrated constant-rate entanglement distillation,
showing that distilling \( n \) Bell pairs
requires approximately \( 3n \) physical Bell pairs as input.
We apply this 3-to-1 ratio to estimate our algorithm's memory requirements.
% as shown in Table~\ref{tab:compare_cost}.
Consequently, although telegate schemes require fewer ancilla qubits than teledata, their substantially higher memory overhead outweighs this advantage.
% While all schemes achieve constant circuit depth,
% both ancilla and memory requirements scale linearly with qubit count.
% From a scalability perspective, although 3 Fanout schemes offer reduced depth,
% their increased ancilla and memory costs make them less favorable.
% Also notice that, in all schemes, the output qubits are not used throughout the entire Fanout, so the subsequent T gates in the circuit can be executed at the same time as the rest of the Fanout. As such, the total depth should be a slightly smaller than the sum. 
Based on these considerations, we recommend the teledata scheme,
as detailed in Table~\ref{tab:compare_cost}b.

% \begin{table}[ht]
%     \centering
%     \resizebox{\columnwidth}{!}{%
%     \begin{tabular}{lllll}
%     \hline
%     Scheme                  & Ancilla & Bell pairs & Depth & Memory est. \\ \hline
%     (a) Telegate, 4 Fanouts (Table~\ref{tab:telegate_cost_4fanouts}) & n       & 2 + 6n     & 95    & 19n + 6 \\
%     % (b) Telegate, 3 Fanouts (Table~\ref{tab:telegate_cost_3fanouts}) & 2n      & 2 + 6n     & 85    & 20n + 6 \\
%     \textbf{(b) Teledata, 4 Fanouts} (Table~\ref{tab:teledata_cost_4fanouts}) & \textbf{2n}      & \textbf{2 + 4n}     & \textbf{91}    & \textbf{14n + 6} \\
%     % (d) Teledata, 3 Fanouts (Table~\ref{tab:teledata_cost_3fanouts}) & 3n      & 2 + 4n     & 81    & 15n + 6 \\
%     (c) Naive Distribution (Sec.~\ref{sec:distributing_rhos}) & $n$      & $n(n+1)-\frac{n}{k}(\frac{n}{k}+1)$     & 91     & $\sim 3n^2$ \\
%     \hline
%     \end{tabular}%
%     }
%     \caption{
%         Cost per QPU for the proposed algorithm.
%         By distillation protocols
%         in~\cite{pattison_fast_2024,ataides_constant-overhead_2025} where
%         about 3 Bell pairs are required to distill 1 Bell pair,
%         we estimate the memory by 3 times of the Bell pairs + ancilla.
%         The bold font indicates the best scheme we recommend.
%     }
%     \label{tab:compare_cost}
% \end{table}

\section{Error analysis}
\label{sec:error_analysis}

\subsection{Circuit-level noise analysis for constant-depth Fanout}
\label{sec:circuit-level-noise-fanout}
Circuits that use Fanout gates can be simulated
by modeling the noisy circuit as 
an ideal Fanout gate followed by a Pauli error sampled from a distribution.
More specifically,
let \( E_i \) denote a possible Pauli error with probability \( p_i \),
and \( U_\text{ideal} \) and \( U_\text{noisy} \) 
denote the ideal and noisy Fanout gate.
Note that Fanout includes mid-circuit measurements
and thus is not a unitary operation,
but becomes a unitary operation after tracing out the ancilla qubits, leaving only the data qubits.
% \JM{``but becomes unitary after tracing out the ancilla qubits, leaving only ..."}\KL{this is better; modified.}
Our goal is to find a distribution \( \set{(E_i, p_i)} \)
such that
$E_i \cdot U_\text{ideal} = U_\text{noisy}
\Longrightarrow E_i = U_\text{noisy}\, \cdot U_\text{ideal}^{-1}.$
Since Fanout is a Clifford circuit, we can use Stim~\cite{Gidney2021stim} %\verb|Stim| 
to simulate it efficiently. 
% In addition, 
% the Fanout circuit includes dynamic logic such as mid-circuit measurements
% and classical feedforward which can be handled by \verb|Stim.TableauSimulator|.
For a given noise level \( p \),
we apply (a) a depolarizing error rate of \( p/10 \) to single-qubit gates,
(b) a depolarizing error rate of \( p \) to two-qubit gates,
and (c) a measurement error rate of \( p \).
We vary the noise level \( p \) and number of targets
for the Fanout circuit in Fig.~\ref{fig:fanout}, using 100,000 shots per simulation.

The top 4 errors and their corresponding probabilities are shown in Table~\ref{tab:fanout_errors}.
Here, for example, ``ZIIIX: 0.05\%'' means a Z error on the control qubit 
and an X error on the last target qubit with probability 0.05\%.
As we can see, the highest probability error 
(``1st Error'' column in Table~\ref{tab:fanout_errors})
is always a Z error on the control qubit.
We hypothesize that this is because
any measurement error on the outcomes that 
the Pauli frame correction depends on 
will lead to an incorrect Pauli frame correction.
As shown in Fig.~\ref{fig:fanout},
there is a Z correction on the control qubit
conditioned on multiple measurement outcomes.
Furthermore, for the same reason,
we observe X errors on the target qubits
since they have X corrections conditioned on measurement outcomes
of ancilla qubits.

\begin{table}[!t]
    \centering
    \resizebox{1.0\columnwidth}{!}{%
    \begin{tabular}{llllll}
        \toprule
        $p_{\text{phy}}$ & \#Trgt & 1st Error & 2nd Error & 3rd Error & 4th Error \\
        \midrule
        0.001 & 4 & ZIIII: 0.35\% & IIIXX: 0.13\% & IXXXX: 0.12\% & IIIIX: 0.05\% \\
0.003 & 4 & ZIIII: 1.01\% & IIIXX: 0.37\% & IXXXX: 0.35\% & IIIIX: 0.15\% \\
0.005 & 4 & ZIIII: 1.64\% & IIIXX: 0.70\% & IXXXX: 0.58\% & IIIIX: 0.22\% \\
0.001 & 6 & ZIIIIII: 0.54\% & IIIXXXX: 0.14\% & IXXXXXX: 0.14\% & IIIIIXX: 0.13\% \\
0.003 & 6 & ZIIIIII: 1.52\% & IIIIIXX: 0.41\% & IIIXXXX: 0.40\% & IXXXXXX: 0.35\% \\
0.005 & 6 & ZIIIIII: 2.46\% & IIIIIXX: 0.63\% & IIIXXXX: 0.60\% & IXXXXXX: 0.56\% \\
0.001 & 8 & ZIIIIIIII: 0.73\% & IIIIIIIXX: 0.15\% & IIIIIXXXX: 0.13\% & IIIXXXXXX: 0.13\% \\
0.003 & 8 & ZIIIIIIII: 2.07\% & IIIXXXXXX: 0.42\% & IIIIIIIXX: 0.41\% & IIIIIXXXX: 0.38\% \\
0.005 & 8 & ZIIIIIIII: 3.27\% & IIIIIIIXX: 0.68\% & IIIXXXXXX: 0.61\% & IIIIIXXXX: 0.61\% \\
    \bottomrule
    \end{tabular}
    }
    \caption{Simulation of Fanout gates with different noise levels and numbers of targets.
    Top 4 errors and their corresponding probabilities are shown for each setting.
    The leftmost matrix corresponds to the error on the control qubit and ``I'' indicates the identity matrix.
    % E.g., ``ZIIIX: 0.05\%'' means a Z error on the control qubit 
    % and an X error on the last target qubit
    % with probability 0.05\%.
    }
    \label{tab:fanout_errors}
\end{table}

\subsection{Circuit-level noise analysis for \cswap{} operations}
\label{sec:circuit-level-noise-cswap}

\begin{figure*}[h]
\centering
\setlength{\tabcolsep}{5pt} % 
    \subfloat[\centering Fidelity of $r$-party GHZ State]{
    \includegraphics[trim={0 0 0 0},width=0.35\linewidth]{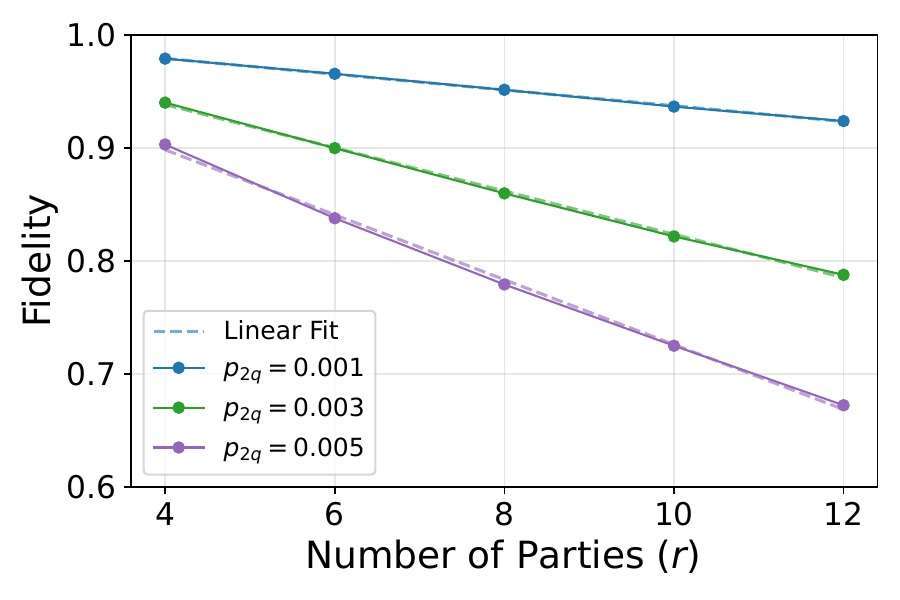}
       \label{fig:ghz-sim-results}
    }
    \subfloat[\centering Classical fidelity of CSWAP via teledata (left) and telegate (right) methods]{
        \includegraphics[trim={0 0 0 0},width=0.65\linewidth]{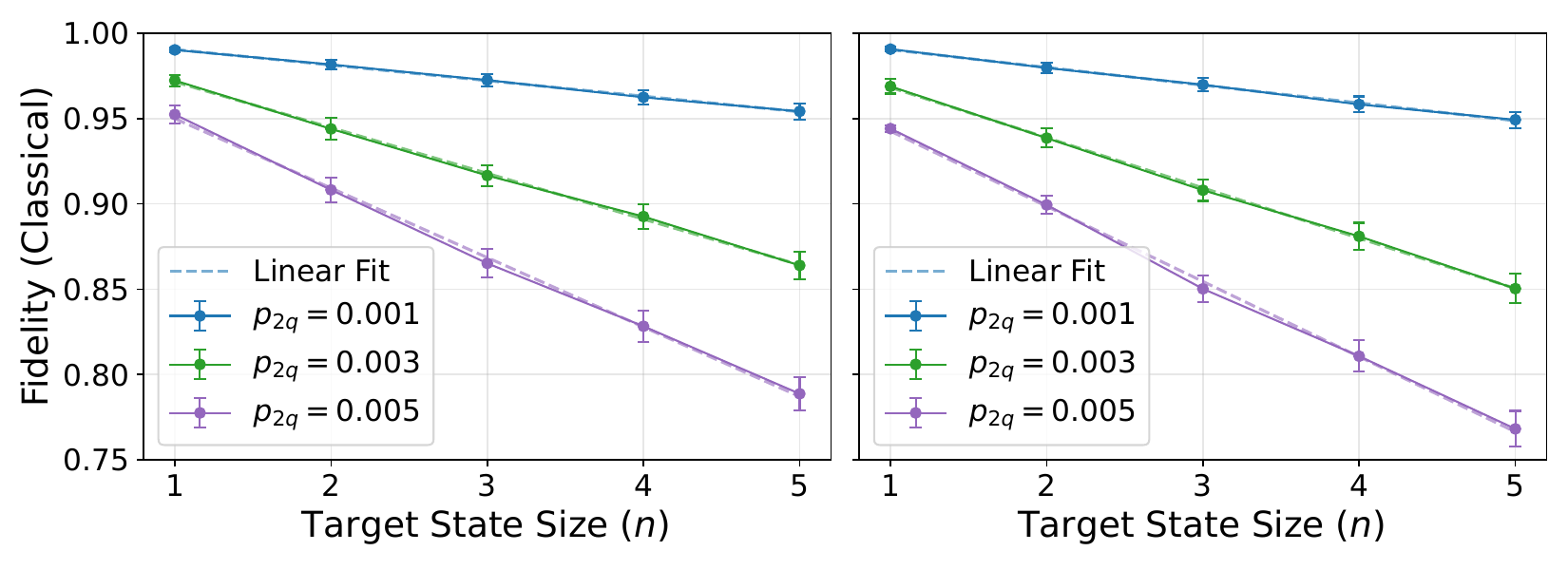}
        \label{fig:cswap-sim-results}
    }
    
\begin{tabular}{cc}

    \subfloat[\centering Overall fidelity estimates for \name{} via teledata (left) and telegate (right) methods]{
    \includegraphics[trim={0 0 0 0}, width=0.65\linewidth]{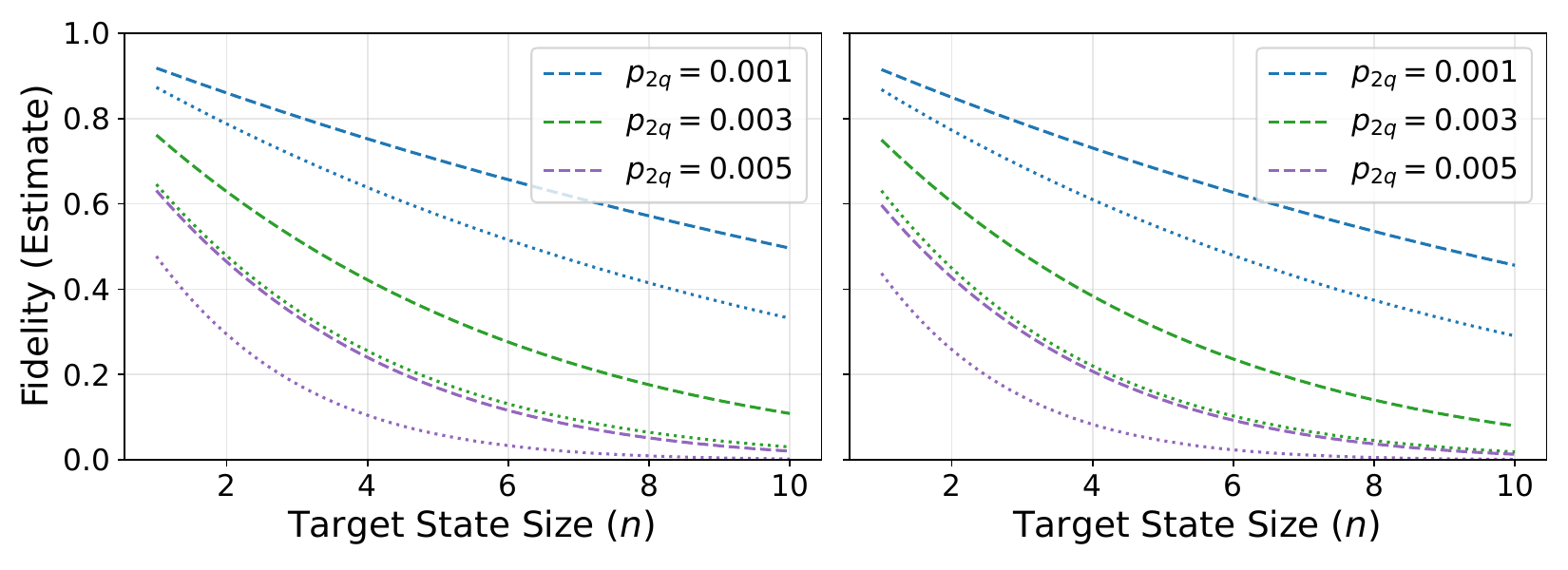}
    \label{fig:overall-sim-results}
    }

 \begin{minipage}[t]{0.35\linewidth}
    \vspace{-12.5em}
    \caption{Simulation results for the two main components of \name{} are shown in (a) and (b). We then combine these in (c) to produce an overall fidelity estimate for \name{}. Here, we plot distributing among $k = 8$ QPUs (dashed line) and $k=12$ QPUs (dotted line) .}
\end{minipage}
    
\end{tabular}

\end{figure*}

For the \cswap{} circuit---whose size makes density-matrix simulation impractical and whose fidelity depends on the input state---we use Qiskit's shot-based simulator to sample classical measurement outcomes~\cite{qiskit2024}. The circuit acts on a Hilbert space of dimension $2^{2n + 1}$. For cases where $2^{2n + 1} \leq 300$, we simulate the circuit exhaustively over all possible pure computational-basis states. When $2^{2n + 1} > 300$, we randomly sample 300 input basis states. In this stage, we only have access to the classical output distributions, so for each input, we compute a classical fidelity, defined as the fraction of measurement outcomes matching the expected noiseless output. We plot the classical fidelity as a function of target state width in Fig.~\ref{fig:cswap-sim-results} for both schemes. 

The simulation results show comparable performance between the two schemes, with the fidelity of the telegate scheme averaging about $0.84\%$ less than that of the teledata scheme. For both designs, fidelity decreases as $n$ increases with sharper drop-offs as $p_{2q}$ increases. 
% The raw data and code for reproducing the simulations can be found in the \href{https://github.com/kunliu7/Distributed-Q-Algo}{GitHub repo}. \KL{In submission, I think we should not use GitHub since it releases information of authors? I suggest saying "We will open-source the code upon acceptance."}

To make the simulations tractable, we reduce the number of ancilla qubits required by blackboxing higher-level primitives such as Fanout, state teleportation, and gate teleportation. Specifically, we execute these operations manually, either via custom unitaries or treating distributed operations as local ones. We then inject noise into the circuit based on Stim simulations of each primitive, as described in Sec.~\ref{sec:circuit-level-noise-fanout} for Fanout errors.
This allows us to simulate larger circuits while preserving realistic noise characteristics. 

\subsection{Circuit-level noise analysis for GHZ state preparation}
\label{sec:circuit-level-noise-ghz}

To evaluate the fidelity of the distributed GHZ state preparation circuit, we again simulate the circuit in Qiskit~\cite{qiskit2024}. For this stage, the circuit is small enough that we may simulate the full density matrix and obtain the noisy output state $\rho$. We then compute the quantum fidelity $\bra{\text{GHZ}}\rho\ket{\text{GHZ}}$ with an ideal GHZ state. Fig.~\ref{fig:ghz-sim-results} shows that the fidelity of the GHZ state produced decreases linearly in $n$.

\subsection{Overall fidelity estimate based on circuit-level noise analysis}
\label{sec:circuit-level-noise-overall}

Simulating the full circuit is computationally prohibitive due to its size.
Instead, we estimate the worst-case fidelity by analyzing the error rates
of individual circuit components.

The total fidelity depends on $n$, the width of \( \rho \), and $k$, the number of QPUs.
We denote \( p_{\text{\cswap{}}}(n) \) as the error rate of the \cswap{} operation
between two \( n \)-qubit states,
and \( p_{\text{GHZ}}(\lceil k/2\rceil) \) as the error rate of generating a \( \lceil k/2\rceil \)-party GHZ state.
In a single multi-party SWAP test, there are two layers of \cswap{} operations
(see Fig.~\ref{fig:PermutationCircuit}).
The total number of \cswap{}s is \( k - 1 \).
Therefore, the total fidelity lower bound for \( n \) and \( k \) is 
$
    (1 - p_{\text{GHZ}}(\lceil k/2\rceil))\, \cdot\, (1 - p_{\text{\cswap{}}}(n))^{k - 1},
$
where \( p_{\text{GHZ}} \) and \( p_{\text{\cswap{}}} \) can be derived from 
Sec.~\ref{sec:circuit-level-noise-cswap} and Sec.~\ref{sec:circuit-level-noise-ghz} respectively. In Fig.~\ref{fig:overall-sim-results}, we plot the estimated fidelity as a function of $n$ for varying values of $k$ and $p_{2q}$. Again, the teledata design slightly outperforms the telegate design and fidelity decreases as $n$, $k$ and $p_{2q}$ increase.

% \begin{figure}
%     \centering
% \includegraphics[trim={0 1cm 0 0}, width=\linewidth]{figs/overall_simulations.pdf}
%     \caption{Overall fidelity estimates for distributed multi-party SWAP tests calculated by combining the simulation results of the individual circuit competences. Here, we plot $k = 8$ (dashed line) and $k=12$ (dotted line).}
%     \label{fig:overall-fid-estimate-plot}
% \end{figure}

\subsection{Network-level noise analysis of Bell pair distribution}

\label{sec:noisy-channels}

\begin{figure}
    \centering

\includegraphics[trim={0 1cm 0 0}, width=\columnwidth]{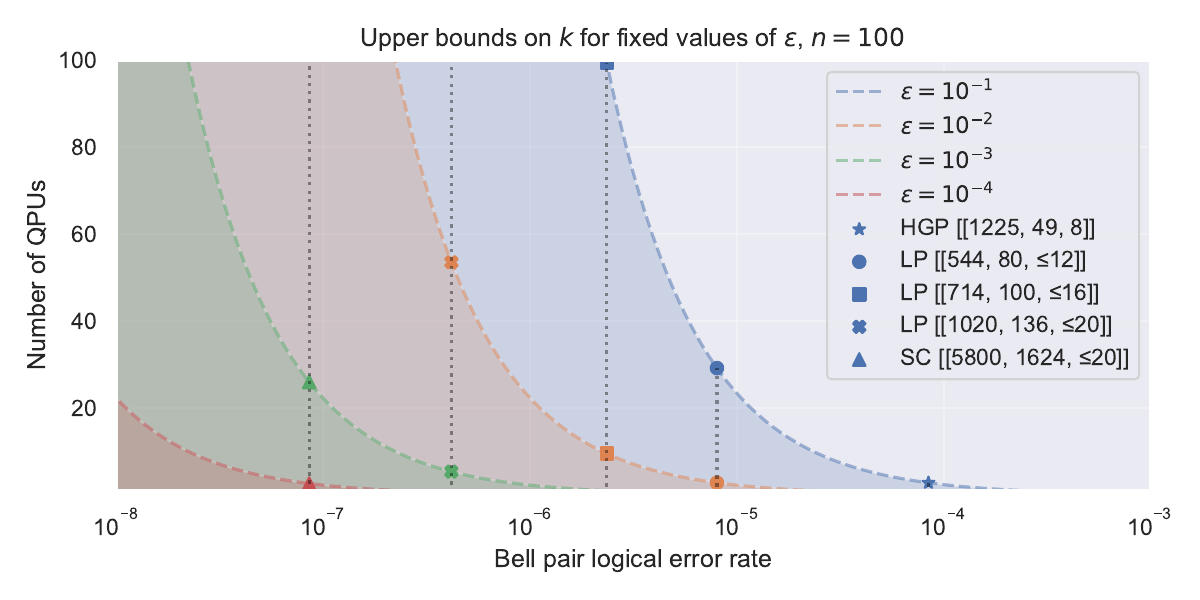}

\caption{Graph of upper bounds on $k$, the number of QPUs among which we can distribute the computation, as a function of $p$, the depolarizing probability for each Bell pair, and error tolerance, $\epsilon$. 
%Each state used in the computation, $\rho_i$ occupies 100 qubits per QPU. 
%Thus, there is a tradeoff between the amount of parallelization we can take advantage of and the amount of noise introduced into the circuit. 
%Here, we show graphs of the telegate method where $k \leq \frac{\log(1- \epsilon)}{(2 + 6n)\log(1 - \frac{3}{4}p)} = \mathcal{O}(\frac{\epsilon}{np})$, as derived in Sec. \ref{sec:noisy-channels}. The teledata method shares the same asymptotic behavior. We also highlight values of $p$ achieved by entanglement distillation schemes shown in \cite{ataides_constant-overhead_2025} and their corresponding upper bounds on $k$ for each value of $\epsilon$. 
% \hzc{Not sure how much this graph adds?}
}
    \label{fig:k_bound}
\end{figure}

In the methods we propose, distributed Bell pairs are essential for both telegate and teledata operations. However, distributing these Bell pairs introduces noise to the circuit. In this section, we assume perfect local gates and use a depolarizing channel with probability $p$ as a model of errors introduced through Bell pair distribution. Under these assumptions, we determine that each of our two protocols for the the distributed multi-party SWAP test has a state fidelity of $F_\text{tot} \geq (1 - \frac{3}{4}p)^{\mathcal{O}(nk)}$. Therefore, to achieve $F_\text{tot} \geq 1 - \epsilon$, we must have $k \leq \mathcal{O}(\frac{\epsilon}{np})$. A graph of these relationships for various values of $\epsilon$ is shown in Fig.~\ref{fig:k_bound}.

When Alice generates a Bell pair and sends one of the qubits to Bob, we model this by a depolarizing channel, $\mathcal{E}$: % https://quantumcomputing.stackexchange.com/questions/12075/how-to-compute-the-tensor-product-of-the-depolarizing-channel-with-the-identity
\begin{equation}
    (I \otimes \mathcal{E})\rho = (1 - p)\rho + p(\rho_A \otimes I/2)
\end{equation}
where $\rho = \ket{\Phi^+}\bra{\Phi^+}$ is the original noiseless Bell pair, and $p$ is the probability that the qubit sent to Bob is completely depolarized by the channel. Thus the state after transmission through the channel is
\begin{equation}
\rho'_{\text{bell}} = (1-p)\ket{\Phi^+}\bra{\Phi^+} + p\frac{1}{4}(I \otimes I) \quad. \label{eqn:noisy-bell}
\end{equation}

We can now simulate the effect of noise on the system by running the CNOT telegate circuit as seen in Fig. \ref{fig:gate-tele} but replacing the ideal Bell pair with $\rho'_{\text{bell}}$. By linearity, this results in a state $(1-p)\rho^{\text{CNOT}}_{\text{pure}} + p\rho^{\text{CNOT}}_{\text{dep}}$, where $\rho^{\text{CNOT}}_{\text{pure}}$ comes from the ideal component and $\rho^{\text{CNOT}}_{\text{dep}}$ comes from the depolarized component. 

We can then compare the quality of this state with the output state produced by the noiseless version by calculating the fidelity $F = (1-p) + p\bra{\Psi_{\text{CNOT}}}\rho^{\text{CNOT}}_\text{dep}\ket{\Psi_{\text{CNOT}}}$.  In  Appendix \ref{sec:cnot-error-bound-appendix}, we show that $F$ achieves a minimum value of $\frac{1}{4}$. Therefore, the fidelity of CNOT teleportation with a noisy Bell pair is bounded as $F_{\text{CNOT}} \geq (1-p) + \frac{p}{4} = 1 - \frac{3p}{4}$.

Similarly, in Appendix \ref{sec:toffoli-error-bound-appendix} we deduce that the depolarized component of the teleported Toffoli circuit achieves a minimum value of $\frac{1}{4}$ as well. We can then repeat the previous calculation to see that $F_{\text{Toffoli}} \geq 1 - \frac{3p}{4}$. Therefore, either type of gate teleportation used in the telegate design will have a fidelity of at least $1 - \frac{3p}{4}$.

Calculating the same quantities for state teleportation gives us $\rho^{\text{teledata}}_\text{dep} = \frac{1}{2}(\ket{0}\bra{0} + \ket{1}\bra{1}) = \frac{1}{2}I$,  and thus 

\begin{equation}
    \bra{\Psi_{\text{teledata}}}\rho^{\text{teledata}}_\text{dep}\ket{\Psi_{\text{teledata}}} = \frac{1}{2}.
\end{equation}
This gives us a fidelity for state teleportation given a noisy Bell pair of $F_{\text{teledata}} \geq (1-p) + \frac{p}{2} = 1 - \frac{p}{2}$.

When applying multiple remote teleoperations, each Bell pair interacts only with the qubits associated with its respective teleoperation. Therefore, the total fidelity of all tele-operations is the product of each of the fidelities of the individual operations.

We begin by finding the total number of tele-operations used in our proposed methods. First, distributed GHZ state preparation requires at most $\mathcal{O}(k)$ teleported CNOTS. In either design, each of the $k$ QPUs performs $\mathcal{O}(n)$ teleoperations, giving us a total of $\mathcal{O}(nk)$ teleoperations across all QPUs. Each operation has fidelity at least $1 - \frac{3}{4}p$ leading to a total fidelity of $F_{\text{tot}} \geq \big(1 - \frac{3}{4}p)^{\mathcal{O}(nk)} \geq 1 - \frac{3}{4}p\cdot\mathcal{O}(nk)$,
where the second inequality follows from Bernoulli's inequality. Thus, to ensure the total fidelity remains above a fixed error tolerance, $F_\text{tot} \geq 1- \epsilon$, we must have $k \leq \mathcal{O}(\frac{\epsilon}{np})$. Put simply, distributing over more QPUs comes at the cost of a lower fidelity, unless Bell pair noise is significantly reduced.
% Alternatively, using only the first inequality in Eq. \ref{eqn:fid-bernoulli-bound} yields a tighter bound of $k \leq \frac{\log(1 - \epsilon)}{(2+6n)\log(1 - \frac{3}{4}p)}$.

% In the teledata design, each of the $k$ QPUs perform two teleported CNOTs followed by $4n$ state teleportations. The corresponding total fidelity is
% \begin{equation}
%     F_{\text{tot}} \geq (1 -\frac{3}{4}p)^{2k}\big(1 - \frac{1}{2}p)^{4nk} \geq \big(1 - \frac{3}{4}p)^{(2+4n)k},
% \end{equation}
% yielding the same asymtotic scaling for the number of QPUs: $
%     k \leq \frac{\log(1 - \epsilon)}{(2+4n)\log(1 - \frac{3}{4}p)} = \mathcal{O}(\frac{\epsilon}{np})$.

Recent methods have demonstrated remote Bell pair preparation via photonic interconnects. These methods have been implemented on trapped-ion hardware with fidelity of 0.970(4) \cite{Saha2025}, and on neutral-atom hardware with post-selected fidelity of 0.987(22) \cite{Ritter_2012}. The entanglement distillation schemes developed in \cite{ataides_constant-overhead_2025} enable us to significantly reduce the logical Bell pair infidelity to below $10^{-6}$ with a lifted product (LP) code that has a rate of roughly $\frac{1}{7}$. Other entanglement distillation schemes \cite{pattison_fast_2024} demonstrate even lower infidelity rates at the cost of additional memory overhead. Even assuming this relatively modest fidelity, with $n=100$ qubits per QPU, the LP code allows for up to $k=5$ QPUs involved in the computation before the infidelity due to Bell pair noise surpasses $\epsilon=10^{-3}$. Fig.~\ref{fig:k_bound} plots several codes from \cite{ataides_constant-overhead_2025} against corresponding values of $k$ for several values of $\epsilon$. As Bell pair preparation techniques and entanglement distillation schemes continue to develop, these values are likely to improve as well.

\section{Applications}
\label{sec:applications}
Our distributed implementation of the multi-party SWAP test retains its broad utility, while extending its applications to distributed quantum computing. Here we highlight several of these applications, including Rényi entropy estimation (Sec.~\ref{sec:renyi}), entanglement spectroscopy (Sec.~\ref{sec:entanglement_spectroscopy}), virtual cooling and distillation (Sec.~\ref{sec:virtual_cooling}), and parallel quantum signal processing (Sec.~\ref{sec:qsp}). 

The core of each of these algorithms is the multi-party SWAP test,  applied to a different states and varying numbers of parties $k$. Consequently, their distributed counterparts inherit the same resource and error analysis as presented in Sec.~\ref{sec:resources} and Sec.~\ref{sec:error_analysis}, including the overall resource cost for each QPU, the total number of Bell pairs, and circuit depth.

\subsection{Distributed Algorithms for Estimating Rényi Entropy}
\label{sec:renyi}

The order-$n$ Rényi entropy of a state $\rho$ is defined as 
\begin{equation}
    S_n(\rho) = \frac{1}{1-n}\log(\tr(\rho^n)) , 
\end{equation}
where the limit $n\rightarrow 1$ corresponds to the von Neumman entropy $S(\rho) = -\tr(\rho \log\rho)$. Like the von Neumann entropy, the R\'enyi entropy quantifies the entanglement and spectral properties of a quantum state. In contrast however, the R\'enyi entropy is often easier to evaluate on a quantum computer.

For integers $n > 1$, the computation of the R\'enyi entropy boils down to estimating the trace of the product $\rho^n$, which is trivially achieved by application of the multi-party SWAP test to $n$ copies of $\rho$: $\tr( W_{\sigma} \rho^{\otimes n}) = \tr(\rho^n)$. Indeed, this simple method of computing the R\'enyi entropy has been widely used to study entanglement in both quantum and classical simulations~\cite{Yirka_2021, Islam_2015, Hastings_2010}. Our distributed version of the multi-party SWAP test thus extends the reach of these methods to distributed quantum computation.

\subsection{Distributed Entanglement Spectroscopy}
\label{sec:entanglement_spectroscopy}
While the R\'enyi entropy evaluated at a single integer $n$ provides a snapshot of a state's entanglement, values taken at several integer orders $n$ can collectively provide information about the spectrum of the underlying state. This task is known as entanglement spectroscopy: given a state $\rho$, determine the eigenvalues of its \textit{entanglement Hamiltonian} $H_E = -\log (\rho)$~\cite{Li_2008}. The entanglement Hamiltonian spectrum provides a measure of topological order~\cite{Pollman_2010, Fidkowski_2010, Yao_2010}, exhibits a correspondence with the spectral gap of a state's parent Hamiltonian~\cite{Cirac_2011, Martyn_2020}, and enables location of quantum critical points~\cite{Calabrese_2008}.

The authors of~\cite{Johri_2017, Suba__2019} introduce a method that uses an elementary mathematical identity, known as the Newton-Girard formula, to extract the spectrum of $H_E$ from values of R\'enyi entropy at several integer orders $n$. Specifically, the Newton-Girard formula re-expresses the characteristic polynomial of the entanglement Hamiltonian as 
\begin{equation}
    \prod_{i} (x-\lambda_i) = \sum_j c_j x^j , 
\end{equation}
where $\lambda_i$ are the eigenvalues of $H_E$, and the coefficients $c_j$ are linear combinations of the integer order R\'enyi entropies (see~\cite{Johri_2017} for the precise expressions). Therefore, by estimating many R\'enyi entropies and then numerically finding the roots of the above polynomial, one obtains an estimate of the entanglement spectrum. Consequently, one can use our distributed multi-party SWAP test iteratively to determine R\'enyi entropies at various orders $n$ and thus estimate the entanglement spectrum. This provides a distributed protocol for performing entanglement spectroscopy.

\subsection{Virtual Cooling and Distillation on a Distributed Quantum Computer}
\label{sec:virtual_cooling}
The multi-party SWAP test can also provide sharper estimates of a target state and/or observables. Given $n$ copies of a state $\rho$ and an observable $O$ acting on the Hilbert space of $\rho$, note that the multi-party SWAP test enables one to extract the trace of $O \rho^n$ as:
\begin{equation}
    \tr(W_\sigma \cdot \rho^{\otimes n} \cdot (O \otimes I^{\otimes (n-1)}) ) = \tr(O \cdot \rho^n)
\end{equation}
Equivalently, this allows for the estimation of expectation values with respect to
a \textit{multiplicative product state} 
\begin{equation}
    \chi \propto \rho^n . 
\end{equation} 
This constitutes a powerful and broadly applicable subroutine.

For instance, if $\rho$ is a thermal state at inverse temperature $\beta$, i.e. $\rho = e^{-\beta H}/Z$, then the multiplicative product state is a thermal state at inverse temperature $n\beta$:
\begin{equation}
    \chi = \frac{\rho^n}{\tr(\rho^n)} = \frac{e^{-n\beta H}}{\tr(e^{-n\beta H})} . 
\end{equation}
This effectively reduces the temperature of the thermal state by a factor of $n$, without directly preparing the lower temperature state. This procedure, appropriately known as ``virtual cooling'', enables extraction of ground state properties from multiple copies of thermal states, and has been experimentally realized in ultracold atoms~\cite{Cotler_2019}.

A similar strategy is useful in the context of error mitigation. Suppose $\rho$ is a noisy mixed state, whose dominant eigenvector is a target pure state $|\psi\rangle$. Then, the multiplicative product state $\chi = \rho^n/\tr(\rho^n)$ rapidly converges to $|\psi\rangle \langle \psi |$ with increasing $n$. This process suppresses noise in $\rho$ and allows for accurate computation of expectation values of $|\psi\rangle$, without ever directly preparing this pure state. This method is known as ``virtual distillation'', and is particularly useful in mitigating errors in noisy quantum computations~\cite{Huggins_2021}.

At the heart of both methods is the ability to take expectation values of the multiplicative product state $\chi = \rho^n/\tr(\rho^n)$, which in both~\cite{Cotler_2019, Huggins_2021} is achieved via the multi-party SWAP test. Thus, our distributed implementation of this subroutine extends the capabilities of virtual cooling and virtual distillation to distributed quantum computing architectures, similar to Ref. \cite{tenzan}.

\subsection{Distributed Quantum Signal Processing}
\label{sec:qsp}
Lastly, we discuss quantum algorithmic applications. Here, quantum signal processing (QSP) has recently emerged as a primitive for designing quantum algorithms~\cite{Low_2016, Low_2017, Low2019hamiltonian}. Its key feature is the ability to implement arbitrary polynomial transformations of an operator, such as a Hamiltonian~\cite{Gilyen_2019_QSVT}. Given this flexibility, QSP has been used to unify previously-disparate constructions of quantum algorithms, while maintaining near-optimal performances~\cite{Martyn_2021_Grand}. 
%Beyond this, the method of QSP has also been generalized to  multivariate polynomials~\cite{Rossi_2022_multivariable, multivariateQSP}, random polynomials~\cite{martyn_2025_halving}, and fruitful connections to nonlinear Fourier analysis~\cite{alexis_2024_quantum, laneve_2025_generalized}. 

In general, constructing a degree-$d$ polynomial transformation through QSP requires a circuit of depth $\mathcal{O}(d)$. Recently,~\cite{martyn_2024_parallel} proposed a method to alleviate this constraint by parallelizing QSP across multiple systems. 
%Broadly speaking, parallel algorithms divide a problem into multiple subproblems, solving them independently (e.g., on separate nodes) and more efficiently than the original problem, and finally recombine these solutions to solve the original problem. 
Here, parallelism is naturally achieved by splitting up QSP polynomials into smaller polynomials, which enables parallel computations at reduced circuit depths.

In more detail,~\cite{martyn_2024_parallel} focuses on the problem of estimating non-linear functions of a quantum state $\rho$: $\tr(F(\rho))$ for some function $F(x)$. In ordinary QSP, this is estimated by approximating $F(x) \approx P(x)$ by a degree-$d$ polynomial, and then realizing $P(\rho)$ through QSP with a circuit of depth $\mathcal{O}(d)$. In contrast, parallel QSP parallelizes this computation over $k$ systems, reducing the depth to $\mathcal{O}(d/k)$. This is achieved by factoring the target polynomial into $k$ degree $\mathcal{O}(d/k)$ factor polynomials: $P(\rho) = \prod_{j=1}^k P_j(\rho)$ (up to some mild constraints, which are addressed in~\cite{martyn_2024_parallel}). By realizing these factor polynomials via QSP and arranging them in parallel, the multi-party SWAP test can be used to extract the trace of their corresponding product, which reproduces $P(\rho)$. Because the factor polynomials  have degree $\mathcal{O}(d/k)$, the requisite circuit depth is $\mathcal{O}(d/k)$.

The essential ingredient in Parallel QSP is the multi-party SWAP test. As such, our adaptation of this ingredient to the distributed setting lifts Parallel QSP to \textit{Distributed QSP}, rendering it compatible with distributed quantum architectures. %We illustrate and compare these protocols in Figure~\ref{fig:pqsp-to-dqsp}. 

\begin{comment}
\begin{figure}
    \centering
    \includegraphics[width=\linewidth]{figs/PQSPtoDQSP.pdf}
    \caption{Comparison of parallel and distributed QSP architectures. The left diagram reproduces the parallel QSP setup from~\cite{martyn_2024_parallel}, while the right illustrates our proposed distributed extension in which QPUs communicate only via classical channels or pre-shared Bell pairs. \JM{If we're tight on space, we could afford to drop this figure} }
    \label{fig:pqsp-to-dqsp}
\end{figure}
\end{comment}

% \section{Related Work}
% \input{sections/06RelatedWork}

% \vspace{-1\baselineskip}
\section{Conclusion and Discussion}
%\textcolor{red}{YL: Conclusion is revised. For John: can you make the following two comments more specific? Or directly edit?}
We have presented a new distributed protocol for the multi-party SWAP test, which constitutes a key algorithmic primitive in quantum computing. At the core of our protocol are two different constructions to realize the two-party controlled-SWAP, both enabled by a constant-depth Fanout scheme achieved through teledata or telegate primitives. We perform a detailed resource analysis of our protocol in terms of ancilla qubits, Bell pair consumption, and total circuit depth. To analyze the performance of our protocol under realistic noisy conditions, we present detailed circuit- and network-level noise analyses that quantify the tradeoff between the number of distributed nodes and both the gate and Bell pair distribution noise. Combining both network error and local computation error, the overall error tolerance of the entire protocol is analyzed as well. % \JM{Clarify this sentence, it's confusing to follow}.
We further show that this construction enables various applications in distributed quantum computing, including R\'enyi entropy estimation, entanglement spectroscopy, error mitigation via virtual cooling/distillation, and parallel QSP.
%By combining this with the recent work on parallel quantum signal processing, we show a fully distributed realization of the parallel QSP for estimating polynomials of reduced density matrices, as well as applications in calculating Renyi entropy and performing entanglement spectroscopy. Moreover, we also showcase the power of our distributed protocol in virtual cooling and virtual distillation, sample-based Hamiltonian simulation, and quantum PCA/quantum ML.

While our work marks significant progress towards distributed quantum algorithms and architectures, there remain many open questions. First, regarding resource estimation, we quantified the resource cost at the logical level with noisy analysis. In the future, a more in-depth analysis including error correction and Bell pair distillation overhead would help determine the physical resource requirements for these distributed protocols. Moreover, on the architecture side, the performance of our protocol could be enhanced by a detailed study of quantum network topology and connectivity, location of the Bell pair generation nodes, and heterogeneity on the quantum channels across different nodes. Our work serves as a foundation for future advancements in optimizing resource allocation, taking into account all of these realistic parameters, and could benefit from the development of new classical simulation methods and programming languages for distributed quantum computing.
% \JM{This isn't quite true. General quantum algorithms cannot be parallelized through parallel QSP, so I wouldn't claim that distributed QSP could serve as a unified framework for distributed QC}
% From an algorithm perspective, QSP and its generalizations can achieve a unified treatment of modern quantum algorithms. This suggests that the distributed QSP algorithm developed in this work can potentially serve as a unified framework for general purpose distributed quantum computation. 
Lastly, to enable a broader class of distributed quantum computing applications, it would be advantageous to generalize the multi-party SWAP test to accommodate more generic multivariate trace estimation protocols, such as estimating sums of several multi-party SWAP test in a single circuit \cite{quek_2024}. 
%\JM{What more general multivariate trace estimation protocols are you referring to? Maybe we could include just one example, i.e. ``such as".} 
Applied to QSP, such capabilities would enable the computation of multivariate polynomials in a distributed fashion, which would be particularly valuable useful for large-scale  quantum simulation in quantum chemistry and materials science~\cite{liu2023bootstrap, alexeev2024quantum}.

\section{Code Availability}
The code for the simulations conducted in this paper is available at \href{https://github.com/kunliu7/Distributed-Q-Algo}{github.com/kunliu7/Distributed-Q-Algo}.

\section{Acknowledgments}
The authors would like to acknowledge Brenda Rubenstein and James Tompkin for helpful discussions, mentorship and feedback.

This research was supported by the U.S. Department of Energy, Office of Science, Advanced Scientific Computing Research under contract number DE-SC0025384 and PNNL’s Quantum Algorithms and Architecture for Domain Science (QuAADS) Laboratory Directed Research and Development (LDRD) Initiative. The Pacific Northwest National Laboratory is operated by Battelle for the U.S. Department of Energy under Contract DE-AC05-76RL01830. This research was also supported by the DARPA MeasQuIT program (HR0011-24-9-0359).

YD acknowledges support by the National Science Foundation (under awards CCF-2312754 and CCF-2338063), the Department of Energy Co-Design Center for Quantum Advantage (C2QA), DARPA (HR-0011-23-3-00019), QuantumCT (under NSF Engines award ITE-2302908), Boehringer Ingelheim, and NSF NQVL-ERASE (under award OSI-2435244). External interest disclosure: YD is a scientific advisor to, and receives consulting fees from Quantum Circuits, Inc.

YL is partially supported by NSF OSI-2531350  (with a subcontract from Duke University).

\bibliographystyle{ACM-Reference-Format}
\bibliography{bibliography}
% \appendix 
\appendix

%%%%%%%%%%%%%%%%%%%%%%%%%%%%%%%%%%%%%%%%%%%%%%%%%%%%
% When adding this appendix to your paper, 
% please remove above part
%%%%%%%%%%%%%%%%%%%%%%%%%%%%%%%%%%%%%%%%%%%%%%%%%%%%

\section{Artifact Appendix}

%%%%%%%%%%%%%%%%%%%%%%%%%%%%%%%%%%%%%%%%%%%%%%%%%%%%%%%%%%%%%%%%%%%%%
\subsection{Abstract}

% {\em Obligatory}
The artifact performs circuit-level simulation to 
estimate the fidelities of different schemes implementing the distributed multi-party SWAP test.
To simulate the circuits efficiently,
we divide the circuit into multiple sub-circuits or procedures and simulate them 
individually to derive their error distributions.
These include GHZ state preparation, Fanout, etc. in Sec.~\ref{sec:error_analysis}.
Then for the overall circuit,
we combine the error distributions of the sub-circuits to estimate the overall fidelity.
The fidelities are further displayed in tables and figures.
The code and data is available at the GitHub repository: \url{https://github.com/kunliu7/Distributed-Q-Algo}.

\subsection{Artifact check-list (meta-information)}

% {\em Obligatory. Use just a few informal keywords in all fields applicable to your artifacts
% and remove the rest. This information is needed to find appropriate reviewers and gradually 
% unify artifact meta information in Digital Libraries.}

{\small
\begin{itemize}
  % \item {\bf Algorithm: }
  % \item {\bf Program: }
  % \item {\bf Compilation: }
  % \item {\bf Transformations: }
  % \item {\bf Binary: }
  % \item {\bf Model: }
  % \item {\bf Data set: }
  \item {\bf Run-time environment: } Mac OS or Linux.
  \item {\bf Hardware: } All the results in this paper can be generated on a MacBook Pro with an M1 chip and 16GB RAM.
  % \item {\bf Run-time state: }
  % \item {\bf Execution: }
  \item {\bf Metrics: } Fidelity of the circuits.
  \item {\bf Output: } Raw data, figures and tables.
  \item {\bf Experiments: } Execute the scripts in README.md.
  \item {\bf How much disk space required (approximately)?: } The repository including all the data and code is less than 40MB.
  % \item {\bf How much time is needed to prepare workflow (approximately)?: }
  \item {\bf How much time is needed to complete experiments (approximately)?: } About 5 hours to generate all the data.
  \item {\bf Publicly available?: } Yes.
  \item {\bf Code licenses (if publicly available)?: } MIT License.
  \item {\bf Data licenses (if publicly available)?: } MIT License.
  % \item {\bf Workflow framework used?: }
  % \item {\bf Archived (provide DOI)?: }
\end{itemize}
}

%%%%%%%%%%%%%%%%%%%%%%%%%%%%%%%%%%%%%%%%%%%%%%%%%%%%%%%%%%%%%%%%%%%%%
\subsection{Description}

\subsubsection{How to access}

% {\em Obligatory}

Clone the repository from Github.

\subsubsection{Hardware dependencies}
All the results in this paper can be generated on a MacBook Pro with an M1 chip and 16GB RAM.

\subsubsection{Software dependencies}
Required Python version >= 3.12. Required Python packages and their versions are listed in \verb|requirements.txt| in the repository.

\subsubsection{Data sets}
None.

\subsubsection{Models}
None.

%%%%%%%%%%%%%%%%%%%%%%%%%%%%%%%%%%%%%%%%%%%%%%%%%%%%%%%%%%%%%%%%%%%%%
\subsection{Installation}

% {\em Obligatory}
See README.md. 
Recommended: create a conda environment with Python version >= 3.12 and install the package locally. This requires installation of at least \href{https://docs.conda.io/projects/conda/en/stable/user-guide/install/index.html}{miniconda}.
If you are using \verb|pipenv|, please make sure the Python version >= 3.12.

\subsection{Basic Test}

We use \verb|pytest| to test the basic functionality of the code.
See \href{https://github.com/kunliu7/Distributed-Q-Algo/tree/main?tab=readme-ov-file#run-tests}{here} at README.md.

%%%%%%%%%%%%%%%%%%%%%%%%%%%%%%%%%%%%%%%%%%%%%%%%%%%%%%%%%%%%%%%%%%%%%
\subsection{Experiment workflow}
None.

%%%%%%%%%%%%%%%%%%%%%%%%%%%%%%%%%%%%%%%%%%%%%%%%%%%%%%%%%%%%%%%%%%%%%
\subsection{Evaluation and expected results}

We separate data generation and result visualization.
Step-by-step instructions are listed in README.md.
In summary,
the top-level script to generate the data is introduced at \href{https://github.com/kunliu7/Distributed-Q-Algo/tree/main?tab=readme-ov-file#how-to-run-the-cswap-circuits-top-level}{here}.
% {\em Obligatory}
In README.md's \href{https://github.com/kunliu7/Distributed-Q-Algo/tree/main?tab=readme-ov-file#visualization}{Visualization} section,
we indicate how each figure in this paper is generated.

%%%%%%%%%%%%%%%%%%%%%%%%%%%%%%%%%%%%%%%%%%%%%%%%%%%%%%%%%%%%%%%%%%%%%
\subsection{Experiment customization}

Users can customize the circuit by modifying the arguments
when executing the scripts.

%%%%%%%%%%%%%%%%%%%%%%%%%%%%%%%%%%%%%%%%%%%%%%%%%%%%%%%%%%%%%%%%%%%%%
\subsection{Notes}
None.

%%%%%%%%%%%%%%%%%%%%%%%%%%%%%%%%%%%%%%%%%%%%%%%%%%%%%%%%%%%%%%%%%%%%%
\subsection{Methodology}

Submission, reviewing and badging methodology:

\begin{itemize}
  \item \url{https://www.acm.org/publications/policies/artifact-review-badging}
  \item \url{http://cTuning.org/ae/submission-20201122.html}
  \item \url{http://cTuning.org/ae/reviewing-20201122.html}
\end{itemize}

%%%%%%%%%%%%%%%%%%%%%%%%%%%%%%%%%%%%%%%%%%%%%%%%%%%%
% When adding this appendix to your paper, 
% please remove below part
%%%%%%%%%%%%%%%%%%%%%%%%%%%%%%%%%%%%%%%%%%%%%%%%%%%%

\section{Calculation of error bounds in Sec. 5}

As described in Sec.~\ref{sec:error_analysis}, we derive a worst-case bound on the overlap between the ideal state and the noisy state after performing each circuit. 

The symbolic calculations in this section were verified using Mathematica, the code for which is publicly available in the GitHub repository.

\subsection{Remote CNOT Gate}\label{sec:cnot-error-bound-appendix}

Consider the two-qubit input state  
\begin{equation*}
    \ket{\varphi} \otimes \ket{\psi} = a_0b_0\ket{00} + a_0b_1\ket{01} + a_1b_0\ket{10} + a_1b_1\ket{11}.
\end{equation*}
Applying an ideal CNOT gate yields 
 \begin{equation*}
      \ket{\Psi_{CNOT}} = a_0 b_0\ket{00} + a_0 b_1\ket{01} 
      + a_1 b_1\ket{10} + a_1 b_0\ket{11}.
 \end{equation*}
If we perform the remote CNOT gate, but replace the Bell pair with a depolarized state, the resulting state can be computed (e.g. using Mathematica) as

\begin{align*}
      \rho^{\text{CNOT}}_{\text{dep}} = \frac{1}{2}\big(|a_0|^2\ket{00}\bra{00}+ |a_0|^2(b_0^*b_1 + b_1^*b_0)\ket{00}\bra{01}\\  |a_1|^2(b_1^*b_0+b_0^*b_1)\ket{01}\bra{00}+
      |a_0|^2 \ket{01}\bra{01}\\ 
+|a_1|^2\ket{10}\bra{10}+|a_1|^2(b_1^*b_0+b_0^*b_1)\ket{10}\bra{11}\\
    +|a_1|^2(b_0^*b_1+b_1^*b_0)\ket{11}\bra{10}+|a_1|^2\ket{11}\bra{11}\big).
\end{align*}

We can then calculate that 

\begin{align*}
    \bra{\Psi_{\text{CNOT}}}\rho^{\text{CNOT}}_{\text{dep}}\ket{\Psi_{\text{CNOT}}} &= \frac{1}{2}(|a_0|^4+|a_1|^4)\\
    & \times (1+2|b_1|^2 -2|b_1|^4 +2\operatorname{Re}((b_0^*)^2b_1^2)),
\end{align*}
which achieves a minimum value of $\frac{1}{4}$ for $\ket{\varphi} = \ket{+}, \ket{\psi} = \ket{1}$.

Since the depolarizing channel yields the ideal Bell pair with probability \(1-p\) and completely depolarizes it with probability \(p\), the fidelity of the teleported CNOT satisfies

\begin{equation}
    F_{\text{CNOT}} \ge (1-p) + \frac{p}{4} = 1 - \frac{3p}{4}.
\end{equation}

\subsection{Remote Toffoli Gate}\label{sec:toffoli-error-bound-appendix}

We now perform the same calculation for the remote Toffoli gate. Let the control qubits be
\(a_0\ket{0}+a_1\ket{1}\) and \(b_0\ket{0}+b_1\ket{1}\), and the target qubit be
\(c_0\ket{0}+c_1\ket{1}\). Then, the ideal output state of the remote Toffoli gate is

\begin{align*}
    \ket{\Psi_{\text{Toffoli}}} &= a_0 b_0c_0\ket{000} + a_0 b_0c_1\ket{001} + a_0 b_1c_0\ket{010}\\
    &+ a_0 b_1c_1\ket{011} + a_1 b_0c_0\ket{100} + a_1 b_1c_1\ket{110}\\
    &+ a_1 b_1c_0\ket{111}.
\end{align*}

If we replace the Bell pair used in the circuit with the depolarized Bell pair, the output density matrix can again be computed explicitly as

\begin{align*}
\rho^{\text{Toffoli}}_{\text{dep}} = \frac{1}{2} \big(|a_0 b_0|^2 \ket{000}\bra{000} +  |a_0 b_0|^2 2\operatorname{Re}(c_1^* c_0) \ket{000}\bra{001} \\
+  |a_0|^2 b_1^* b_0 \ket{000}\bra{010} +  |a_0|^2 b_1^* b_0 2\operatorname{Re}(c_1^* c_0)\ket{000}\bra{011}\\
+  |b_0|^2 a_1^* a_0 \ket{000}\bra{100} +  |b_0|^2 a_1^* a_0 2\operatorname{Re}(c_1^* c_0) \ket{000}\bra{101} \\
+  |a_0 b_0|^2 2\operatorname{Re}(c_1^* c_0) \ket{001}\bra{000} +  |a_0 b_0|^2 \ket{001}\bra{001} \\
+  |a_0|^2 b_1^* b_0 2\operatorname{Re}(c_1^* c_0) \ket{001}\bra{010} +  |a_0|^2 b_1^* b_0 \ket{001}\bra{011} \\
+  |b_0|^2 a_1^* a_0 2\operatorname{Re}(c_1^* c_0) \ket{001}\bra{100} +  |b_0|^2 a_1^* a_0 \ket{001}\bra{101} \\
+  |a_0|^2 b_0^* b_1 \ket{010}\bra{000} +  |a_0|^2 b_0^* b_1 2\operatorname{Re}(c_1^* c_0) \ket{010}\bra{001} \\
+  |a_0 b_1|^2 \ket{010}\bra{010} +  |a_0 b_1|^2 2\operatorname{Re}(c_1^* c_0) \ket{010}\bra{011} \\
+  a_1^* b_0^* a_0 b_1 \ket{010}\bra{100} +  a_1^* b_0^* a_0 b_1 2\operatorname{Re}(c_1^* c_0) \ket{010}\bra{101} \\
+  |a_0|^2 b_0^* b_1 2\operatorname{Re}(c_1^* c_0) \ket{011}\bra{000} +  |a_0|^2 b_0^* b_1 \ket{011}\bra{001} \\
+  |a_0 b_1|^2 2\operatorname{Re}(c_1^* c_0) \ket{011}\bra{010} +  |a_0 b_1|^2 \ket{011}\bra{011} \\
+  a_1^* b_0^* a_0 b_1 2\operatorname{Re}(c_1^* c_0) \ket{011}\bra{100} +  a_1^* b_0^* a_0 b_1 \ket{011}\bra{101} \\
+  |b_0|^2 a_0^* a_1 \ket{100}\bra{000} +  |b_0|^2 a_0^* a_1 2\operatorname{Re}(c_1^* c_0) \ket{100}\bra{001} \\
+  a_0^* b_1^* a_1 b_0 \ket{100}\bra{010} +  a_0^* b_1^* a_1 b_0 2\operatorname{Re}(c_1^* c_0) \ket{100}\bra{011} \\
+  |a_1 b_0|^2 \ket{100}\bra{100} +  |a_1 b_0|^2 2\operatorname{Re}(c_1^* c_0) \ket{100}\bra{101} \\
+  |b_0|^2 a_0^* a_1 2\operatorname{Re}(c_1^* c_0) \ket{101}\bra{000} +  |b_0|^2 a_0^* a_1 \ket{101}\bra{001} \\
+  a_0^* b_1^* a_1 b_0 2\operatorname{Re}(c_1^* c_0) \ket{101}\bra{010} +  a_0^* b_1^* a_1 b_0 \ket{101}\bra{011} \\
+  |a_1 b_0|^2 2\operatorname{Re}(c_1^* c_0) \ket{101}\bra{100} +  |a_1 b_0|^2 \ket{101}\bra{101} \\
+  |a_1 b_1|^2 \ket{110}\bra{110} +  |a_1 b_1|^2 2\operatorname{Re}(c_1^* c_0) \ket{110}\bra{111} \\
+  |a_1 b_1|^2 2\operatorname{Re}(c_1^* c_0) \ket{111}\bra{110} +  |a_1 b_1|^2\big) \ket{111}\bra{111}
\end{align*}

Plugging in the ideal Toffoli output state \(\ket{\Psi_{\text{Toffoli}}}\), we obtain

\begin{align*}
    \bra{\Psi_{\text{Toffoli}}}\rho_{\text{Toffoli}}^{\text{dep}}\ket{\Psi_{\text{Toffoli}}} &=\frac{1}{2}(1 - 2|a_1|^2|b_1|^2
    + 2|a_1|^4|b_1|^4)\\
    &\times (1+2|c_1|^2 - 2|c_1|^4 +  2\operatorname{Re}((c_0^*)^2c_1^2))
\end{align*}
which achieves a minimum value of $\frac{1}{4}$ for $|a_1||b_1| = \frac{1}{\sqrt{2}}, c_0 = 0, c_1 = 1$. Therefore, we achieve a fidelity
\begin{equation}
    F_{\text{Toffoli}} \geq (1-p) + \frac{p}{4} = 1 - \frac{3p}{4}.
\end{equation}

\end{document}